\documentclass{jfm}

\usepackage[T1]{fontenc}
\usepackage[english]{babel}															
\usepackage{amsmath,amsfonts} 
\usepackage{graphicx}	
\usepackage{caption}
\usepackage{dcolumn}
\usepackage{natbib}
\usepackage{bm}
\usepackage{rotating}
\usepackage{textcomp}
\usepackage{tikz}
\usepackage{textcomp}
\usepackage{datetime}
\usepackage{caption}
\usepackage{xspace}
\usepackage{fancyhdr}
\usepackage{mathrsfs}
\usepackage{listings}
\usepackage{subfigure}
\usepackage{multirow,bigstrut}
\usepackage[draft]{changes}
\usepackage{lipsum}

\ifCUPmtlplainloaded \else
  \checkfont{eurm10}
  \iffontfound
    \IfFileExists{upmath.sty}
      {\typeout{^^JFound AMS Euler Roman fonts on the system,
                   using the 'upmath' package.^^J}%
       \usepackage{upmath}}
      {\typeout{^^JFound AMS Euler Roman fonts on the system, but you
                   dont seem to have the}%
       \typeout{'upmath' package installed. JFM.cls can take advantage
                 of these fonts,^^Jif you use 'upmath' package.^^J}%
      }
  \else
  \fi
\fi

\ifCUPmtlplainloaded \else
  \checkfont{msam10}
  \iffontfound
    \IfFileExists{amssymb.sty}
      {\typeout{^^JFound AMS Symbol fonts on the system, using the
                'amssymb' package.^^J}%
       \usepackage{amssymb}%
         \let\leq=\leqslant
         
      }{}
  \fi
\fi

\ifCUPmtlplainloaded \else
  \IfFileExists{amsbsy.sty}
    {\typeout{^^JFound the 'amsbsy' package on the system, using it.^^J}%
     \usepackage{amsbsy}}
    {\providecommand\boldsymbol[1]{\mbox{\boldmath $##1$}}}
\fi

\newcommand\Be{\mathrm{\textit{Be}}}  
\newcommand\Sh{\mathrm{\textit{Sh}}}   

\newcommand\Da{\mathrm{\textit{Da}}}   
\newcommand\Dc{\mathrm{\textit{Dc}}}   

\begin{document}

\shorttitle{Membrane morphology and topology} 
\shortauthor{B Ling and I Battiato} 

\title{Rough or  Wiggly? Membrane Topology and Morphology for Fouling Control}

\author
 {
	Bowen Ling\aff{1}
  \and 
  Ilenia Battiato\aff{1}
	\corresp{\email{ibattiat@stanford.edu}},
  }

\affiliation
{
\aff{1}
Energy Resources Engineering, Stanford University, Stanford, CA, USA
}

\maketitle

\begin{abstract}
Reverse Osmosis Membrane (ROM) filtration systems are widely applied in wastewater recovery, seawater desalination, landfill water treatment, \etc During filtration, the system performance is dramatically affected by membrane fouling which causes a significant decrease in permeate flux as well as an increase in the energy input required to operate the system. Design and optimization of ROM filtration systems aim at reducing membrane fouling by studying the coupling between membrane structure, local flow field, local solute concentration  and foulant adsorption patterns. Yet, current studies focus exclusively on oversimplified steady-state models that ignore any dynamic coupling between the fluid dynamics and the transport through the membrane, while membrane design still proceeds through trials and errors. In this work, we develop a model that couples  the transient Navier-Stokes and the Advection-Diffusion-Equations, as well as an adsorption-desorption equation for the foulant accumulation, and we validate it against unsteady measurements of permeate flux as well as steady-state spatial fouling patterns. Furthermore, we analytically show that, for a straight channel, a universal scaling relationship exists between the Sherwood and Bejam numbers, i.e. the dimensionless permeate flux through the membrane and the pressure drop along the channel, respectively. We then generalize this result to membranes subject to morphological and/or topological modifications, i.e., whose shape (wiggliness) or surface roughness is altered from the rectangular and flat reference case. We  demonstrate that universal scaling behavior can be identified through the definition of a modified Reynolds number, $\Rey^\star$, that accounts for the additional length scales introduced by the membrane modifications, and a membrane performance index, $\xi$, which represents an aggregate efficiency measure with respect to both clean permeate flux and energy input required to operate the system. Our numerical simulations demonstrate that `wiggly' membranes outperforms  `rough' membranes for smaller values of $\Rey^\star$, while the trend is reversed at higher $\Rey^\star$. To the best of our knowledge, the proposed approach is the first able to quantitatively investigate, optimize and guide the design of both morphologically and topologically altered membranes under the same framework, while providing  insights on the physical mechanisms controlling the overall system performance.


\end{abstract}

\section{Introduction}
Reverse Osmosis Membrane (ROM) filtration systems are utilized in wastewater recovery \citep{shannon2008science, greenlee2009reverse, benito2002reverse, rahardianto2010accelerated, mccool2010feasibility, rahardianto2008reverse, cath2005membrane}, seawater desalination \citep{elimelech2011future, fritzmann2007state, matin2011biofouling}, landfill water treatment \citep{chianese1999treatment, peters1998purification}, etc. Typically, ROMs perform one of the final stages of water treatment and are designed to filter ions or soluble substances. Bio-active films \citep{bucs2016biofouling} and porous materials \citep{rahardianto2006diagnostic, shih2005morphometric} are    two of the  media frequently used as  separation membranes. The selective membrane only allows de-mineralized/deionized water to penetrate, and forms a physical boundary between the purified water flux (i.e. the permeate flux), collected on the draw side of the membrane,  and the  pre-treatment (feed) water. High pressure is applied and maintained on the concentrated (feed) side to drive the permeate flux of treated water. As any filtration process, ROMs' performance is largely impacted by fouling. 
The filtered solute (mineral or ion) accumulates on the membrane surface by creating a blockage.  Fouling is the primary process affecting filtration performance since it  (i) reduces the clean water permeate flux and (ii) increases the energy (i.e. the driving pressure drop) required to generate a unitary permeate flux.

While the type of foulant greatly depends on solute and membrane properties, its impact on ROMs performance and operation costs is similar. In bio-active membranes, microbial growth is the primary cause of  fouling \citep{bucs2016biofouling}. Instead,  perm-selectivity of porous ROM membranes to the solvent (e.g. clean water) leads to a localized increase of solute concentration on the feed side, also known as concentration polarization (CP) \citep{brian1965concentration, sablani2001concentration, mccutcheon2006influence, kim2005modeling, jonsson1977concentration}. As a result, solute precipitates from the solution and accumulates or crystallizes on the membrane surface \citep{rahardianto2006diagnostic, shih2005morphometric}. 

Beside surface chemistry, the fouling propensity of a membrane depends greatly on its surface topological properties, such as roughness, and its morphology (or shape). Literature and experience show that modification of membrane surfaces with chemical coatings can be effective but not sufficient for controlling membrane fouling. The discovery that sub-micron patterning of a membrane surface can improve its fouling resistance provides an orthogonal membrane design parameter \citep{maruf2013use,maruf-2013-Influence,maruf-2014-fabrication}.  As a result, different mechanisms at vastly different scales have been proposed to control fouling  \citep{zhang-2016-antifouling}: (i) modifications of membrane/separator   morphology  at the system scale ($\sim \mbox{cm}$) \citep{ma2006numerical,guillen2009modeling,suwarno2012impact,xie2014hydrodynamics,sanaei2017flow}; (ii) modifications of the membrane topology  at the micro-scale ($\sim \mbox{mm}-\mu\mbox{m}$) \citep{ba2010using, elimelech1997role, vrijenhoek2001influence, kang2007novel, battiato-2010-elastic,bowen2000atomic,ladner2012functionalized,ling2016dispersion,maruf2013use,maruf-2013-Influence,maruf-2014-fabrication,Battiato-2014-Effective}; and (iii) chemical or surface treatment to alter the interaction force between  the membrane and the foulant at the nano-scale ($\sim \mbox{nm}$) \citep{sanaei2016flow,kang2007study}.

Attempts to control fouling in ROMs commercial systems have been primarily limited to the inclusion of spacers transverse to the main flow direction. These have showed limited success since they fail to sufficiently perturb the flow field in proximity of the membrane boundary, where fouling is localized. Furthermore, spacers increase the flow resistance, i.e. the energy input to sustain a given pressure drop. More promising has been the use of micro/nano-patterns embossed above the membrane. These are able to more effectively perturb the flow locally, and only mildly impact the overall dissipations of the system. Surface treatment can effectively modify interaction properties, for instance  wettability, roughness, and molecular attraction between the membrane and the foulant, however such modifications generally undergo irreversible degradation during filtration. 


 
Despite a number of studies have experimentally or analytically demonstrated the impact  of morphological and topological alteration on, e.g., solute dispersion and fouling at the system (macro-) scale \citep{maruf2013use,battiato-2010-elastic,griffiths2013control,BattiatoRubol,Rubol2016,ling2016dispersion,Ling-2018-hydrodynamics},  ROMs systems are still primarily optimized by trail and error. This is due to the lack of quantitative understanding of  the impact of morphological and/or topological modifications on membrane fouling at prescribed  operating conditions.  While (semi-)analytical methods can provide general guiding principles and basic process understanding for highly idealized systems \citep{battiato-2012-self,sanaei2016flow,ling2016dispersion,kang2007study,sanaei2017flow}, their direct applicability  to optimization and design of real systems is questionable.  On the other hand, laboratory  experimentation of promising designs, and their consequent optimization, may be prohibitively expensive. 
A number of 2D and 3D numerical simulators have been developed to study fouling  \citep{lyster2007numerical,xie2014hydrodynamics,park2013numerical,bucs2014effect,bucs2016biofouling,Kang-2017-Origin,lyster2007numerical,park2013numerical,xie2014hydrodynamics,bucs2014effect,park2013numerical,bhattacharyya1990prediction}. Yet, they generally do not account for unsteady effects and the coupling between hydrodynamics and membrane fouling, which dynamically alters the permeate flux distribution, i.e. the boundary condition for the flow field \citep{lyster2007numerical}. The flow field is influenced by the local blockage of the membrane, which further impacts the  distribution of the solute in the bulk, its concentration polarization and, finally, fouling formation. A number of models have been proposed to represent different mechanisms leading to fouling (e.g. standard and complete blocking, cake filtration, etc.), and its impact on system scale performance, quantified by, e.g., energy input and permeate flux \citep{Griffith-2014-combined,Griffith-2016-designing,sanaei2017flow}. However, the fouling mechanism is generally not dynamically coupled to the local hydrodynamics, and their dynamic feedbacks on flux reduction is not accounted for. Attempts to incorporate unsteady effects  included the definition of time-dependent absorption  functions to model fouling growth \citep{bucs2016biofouling,bucs2014effect}: although the growth function is time-dependent, the governing equations for flow and transport remain the steady-state.   An explicit treatment of the dynamic coupling between bulk transport, surface fouling and hydrodynamics is necessary to elucidate the mechanisms that control (i) the onset of fouling, (ii) the development of a stable fouling pattern and (iii) the dynamic flux reduction as a result of clogging.



In this work, we develop a three dimensional model and computational framework to study fouling spatio-temporal evolution which captures  (i)  the two-way coupling between  bulk concentration, flow velocity  and  foulant accumulation on the membrane surface,  (ii) the relationship between concentration polarization close to the membrane surface and fouling on the membrane, and (iii) the   initiation and development of the foulant spatial pattern. Such a framework allows us to quantitatively investigate the impact of surface topology (i.e. roughness) and morphology (i.e. wiggliness, shape) on fouling, and to identify  dynamical conditions under which such alterations are warranted.

The paper is organized as follows. In Section \ref{sec:model}, we introduce the model in dimensional (\S\ref{sec:dimensional}) and dimensionless (\S\ref{sec:dimensionless}) form, and derive universal scaling laws in the long time limit for a rectangular flat membrane (\S\ref{sec:scaling}). In Section \ref{sec:numerical}, we first  perform a convergence study (\S\ref{sec:convergence}) and then validate the model against  unsteady permeate flux measurements and steady-state fouling patterns (\S\ref{sec:validation}). Section \ref{sec:synthetic} investigates the impact of morphological and  topological modifications of the membrane shape (i.e. wiggliness) and surface (i.e. roughness) on both foulant accumulation, clean water permeate flux and operating pressure drop. We focus on 18 membrane designs which include 9 purely morphological (M-), 6 purely topological (T-) and 3 hybrid (H-) designs  which include both topological and morphological modifications (\S\ref{sec:numerical_modified}). We then introduce new scaling variables (\S\ref{sec:scaling_variable}) and derive scaling laws (\S\ref{sec:scaling_laws}) valid for all designs. We finally show how the previous framework can be used for membrane shape and surface optimization  (\S\ref{sec:optimization}). We provide concluding remarks in Section~\ref{sec:conclusion}.


\section{Model}\label{sec:model}
\subsection{Governing Equations}\label{sec:dimensional}

\begin{figure}
  \centerline{\includegraphics[width=0.7\textwidth]{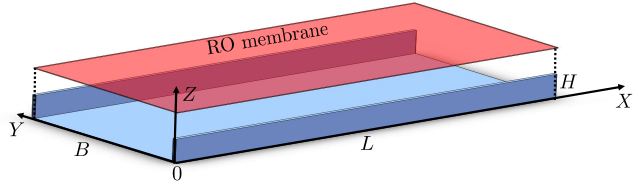}}
  \caption{Three-dimensional sketch of the computational domain, where the red boundary represents the RO membrane and the blue boundaries are the  flow channel walls.}
\label{fig:Domain}
\end{figure}

We consider a pressure-driven flow in a channel of length $L$ and rectangular cross-section (in the $(Y,Z)$-plane) whose top side, located at $Z=H$, consists of a flat RO membrane lying in the $(X,Y)$-plane, parallel to the mean flow. The clean water is cross-filtered  from the feed solution, conveyed to the membrane through the flow channel, as the membrane is permeable to water molecules only and impermeable to the solute dissolved in the feed. The concentrated solution (feed) enters the channel from the inlet section located at $X=0$ and exits the domain at $X=L$. Solute rejection by the membrane (\emph{aka} membrane permselectivity) leads to the emergence of  local solute concentration gradients  in the feed at the membrane/solution interface and to subsequent accumulation of foulant on the membrane, in the interior of the computational domain ($0<Z\leq H$). A schematics of the domain is shown in Figure~\ref{fig:Domain}. 

We focus on fouling accumulation as a function of both time and space. The filtration process is described by a set of coupled transient equations for the velocity field $\mathbf U$,  the bulk concentration $C_{\mbox{\tiny{b}}}$ of solute within the feed solution and the foulant surface concentration $C_{\mbox{\tiny{s}}}$ on the RO membrane. The flow field $\mathbf  U (\mathbf X, T)=(U, V, W)$ satisfies the transient incompressible Navier-Stokes and continuity equations 
\begin{subequations}\label{eq:flow}
\begin{align} 
&\frac{\partial \mathbf{U}}{\partial T}+(\mathbf{U}\cdot\nabla)\mathbf{U}+\nabla \hat{P}=\nabla \cdot(\nu \nabla\mathbf{U}),\label{eqn:NV} \\
&\nabla\cdot \mathbf{U}=0, \label{eqn:continuity}
\end{align}
\end{subequations}
where $\hat{P}$ [L$^2$T$^{-2}$] is a rescaled pressure  and is defined as 
\begin{align}
\hat{P}=\dfrac{P^\star}{\rho}
\end{align} 
with $P^\star$ fluid pressure,  and $\nu$ and $\rho$  the kinematic viscosity and density of the bulk solution, respectively. Gravitational effects are neglected. Equations \eqref{eq:flow} are subject to inlet, outlet, cross-flow velocity and no-slip boundary conditions at the inlet, outlet, on the RO membrane and the three impermeable walls, respectively,
\begin{subequations}\label{bc:flow}
\begin{align}
&\mathbf{U}=(U_{\mbox{\tiny{in}}},0,0), \quad  \mathbf n\cdot \nabla \hat{P}=0 \quad \mbox{for} \quad \mathbf X=(0,Y,Z),\label{bc:inlet}\\
&\mathbf n\cdot \nabla \mathbf{U}=0, \quad \quad \hat{P}=P_{\mbox{\tiny{out}}} \quad \mbox{for} \quad \mathbf X=(B,Y,Z),\label{bc:outlet}\\
&\mathbf{U}=(0,0,W_{\mbox{\tiny{H}}}), \quad  \mathbf n\cdot \nabla \hat{P}=0 \quad \mbox{for} \quad \mathbf X=(X,Y,H),\label{bc:RO}\\
&\mathbf{U}=(0,0,0), \quad  \mathbf n\cdot \nabla \hat{P}=0 \quad \mbox{for} \quad Y=\{0,B\}, \quad \mbox{or} \quad Z=0, \label{bc:walls}
\end{align}
\end{subequations}
where  $W_{\mbox{\tiny{H}}}$ is the local permeate flux through the membrane. It is defined as the difference between $W_{\mbox{\tiny{m}}}$, the clean water flux, i.e. the membrane flux in absence of fouling, and  $W_{\mbox{\tiny{f}}}$, the flux reduction due to foulant accumulation, i.e.
\begin{equation}  \label{BC_membraneV_0}
W_{\mbox{\tiny{H}}}=W_{\mbox{\tiny{m}}}-W_{\mbox{\tiny{f}}}.
\end{equation}
In \eqref{BC_membraneV_0},  $W_{\mbox{\tiny{m}}}$  and $W_{\mbox{\tiny{f}}}$ are defined as follows
\begin{equation}  
W_{\mbox{\tiny{m}}}=\frac{k_{\mbox{\tiny{m}}}}{\nu B} \delta \hat{P},
\end{equation}
with $k_{\mbox{\tiny{m}}}$ [$\mbox{mD}=9.869\times10^{ - 16} \mbox{m}^2$] the membrane permeability, and $\delta \hat{P}$ the local pressure head drop across the membrane,
\begin{align} \label{eqn:dp}
\delta \hat{P}=\hat{P}(X,Y,Z=H^-)-P_{\mbox{\tiny{amb}}}, 
\end{align}
where $P_{\mbox{\tiny{amb}}}=0$ is the ambient pressure. Although various methods have been proposed to model the flux reduction  $W_{\mbox{\tiny{f}}}$ due to fouling, two  approaches are generally used:   the flux reduction is represented either (i) as a function of  the concentration polarization (or concentration in close proximity of the membrane surface) \citep{lyster2007numerical,lee1981membranes,sagiv2014analysis},  or (ii) by postulating a functional relationship between the foulant and membrane resistance \citep{sanaei2016flow,sanaei2017flow,griffiths2013control}.  The first approach is based on the assumption that the fouling and concentration polarization (CP) have the same (or similar) impact on the flow field. However, unlike CP, which vanishes when pressure is released, some foulant may irreversibly precipitate on the membrane  \citep{xie2014hydrodynamics, shih2005morphometric}. The second approach is based on modeling the foulant as an additional resistance to the membrane: in such models, the relationship between foulant-induced resistance  and flux reduction  is postulated and, very often, the effective properties in such relationships (e.g., permeability of the foulant or the attraction coefficient \citep{sanaei2017flow}) are difficult to  determine experimentally.

In this work, instead  we model flux reduction due to foulant accumulation as
\begin{equation} \label{eq:accumulation}
W_{\mbox{\tiny{f}}}=A_{\mbox{\tiny{f}}} \left(C_{\mbox{\tiny{s}}}-C_{\mbox{\tiny{b}}}\right),
\end{equation}
where $A_{\mbox{\tiny{f}}}$ is a constant, and $C_{\mbox{\tiny{s}}}$ [-] and $C_{\mbox{\tiny{b}}}$ [-] are the foulant dimensionless surface and bulk  concentrations defined as
\begin{equation}  
C_{\mbox{\tiny{s}}}=\frac{\hat{C}_{\mbox{\tiny{s}}}}{C_0\cdot B},
\end{equation}
and
\begin{equation}  
C_{\mbox{\tiny{b}}}=\frac{\hat{C}_{\mbox{\tiny{b}}}}{C_0},
\end{equation}
where $\hat{C}_{\mbox{\tiny{s}}}$ [mol/m$^2$] and $\hat{C}_{\mbox{\tiny{b}}}$ [mol/m$^3$] are the corresponding dimensional concentrations,  and $C_0$ is the reference bulk concentration (usually taken as the inlet concentration). The expression $A_{\mbox{\tiny{f}}}(C_{\mbox{\tiny{s}}}-C_{\mbox{\tiny{b}}})^n$ is widely used for studying crystallization kinetics \citep{shih2005morphometric,lee2000effect,cetin2001kinetics,brusilovsky1992flux,sheikholeslami2003kinetics}, similar to those occurring for certain foulants, e.g. gypsum, with an exponent $n$ ranging between 1 and 2. If we normalize the constant $A_{\mbox{\tiny{f}}}$ by the membrane permeability, and define
\begin{equation}  
\hat{A}_0=\frac{A_{\mbox{\tiny{f}}}}{ k_{\mbox{\tiny{m}}}/\left(\nu B\right)},
\end{equation}
then \eqref{BC_membraneV_0} can be written as
\begin{equation} \label{BC_membraneV}
W_{\mbox{\tiny{H}}}=\frac{k_{\mbox{\tiny{m}}}}{\nu B} \left[ \delta \hat{P}- \hat{A}_0\left(C_{\mbox{\tiny{s}}}-C_{\mbox{\tiny{b}}}\right)\right],
\end{equation}
which describes how fouling affects the decrease in permeate flux dynamically:  as the foulant surface concentration, $C_{\mbox{\tiny{s}}}$, increases, the local permeate velocity $W_{\mbox{\tiny{H}}}$ and flux will dynamically decrease. In this context, $\hat A_0\left(C_{\mbox{\tiny{s}}}-C_{\mbox{\tiny{b}}}\right)$ can be thought of as a loss of effective pressure drop across the membrane due to foulant deposition. We emphasize that the boundary condition \eqref{BC_membraneV} allows one to (i) distinguish CP from foulant accumulation, (ii) fully couple the flow field, concentration  field and the foulant deposition while capturing its effect on permeate flux dynamically, and (iii) link  flux reduction  with  foulant accumulation/precipitation on the membrane. 

The solute bulk concentration $C_{\mbox{\tiny{b}}}$ satisfies a classic advection-diffusion equation
\begin{align} \label{eqn:Cb}
\frac{\partial C_{\mbox{\tiny{b}}}}{\partial T}+\mathbf{u}\cdot \nabla C_{\mbox{\tiny{b}}}-D\nabla^2 C_{\mbox{\tiny{b}}}=0,
\end{align}
subject to a flux balancing boundary condition on the RO membrane \citep{lyster2007numerical}
\begin{equation}\label{BC_membraneC}
D\frac{\partial C_{\mbox{\tiny{b}}}}{\partial Z}=R_i W_{\mbox{\tiny{H}}} C_{\mbox{\tiny{b}}} \quad \mbox{at} \quad Z=H,
\end{equation}
where $D$ is the molecular diffusion coefficient of the bulk solute,  $W_{\mbox{\tiny{H}}}$ is given by \eqref{BC_membraneV}, and $R_i$ is the intrinsic membrane rejection rate \citep{lyster2007numerical}. In this study, we set $R_i=100\%$.  Furthermore, Eq. \eqref{eqn:Cb} is subject to no-flux boundary conditions at the outlet and on the channel solid walls, i.e. $\mathbf n\cdot \nabla C_{\mbox{\tiny{b}}}=0$ at $Z=0$, $Y=\{0,H\}$ and $X=L$, and a Dirichlet boundary condition at the inlet, $C_{\mbox{\tiny{b}}}=1$ at $X=0$. The surface concentration of the foulant $C_{\mbox{\tiny{s}}}$ satisfies a transient adsorption-desorption  equation \citep{jones2000protein}.  
\begin{align} \label{eqn:Cs}
\frac{\partial C_{\mbox{\tiny{s}}}}{\partial T}=K_1\cdot(C_{s,\mbox{\tiny{max}}}-C_{\mbox{\tiny{s}}})\cdot C_{\mbox{\tiny{b}}}-K_2 C_{\mbox{\tiny{s}}}, 
\end{align}
where $K_1$, $K_2$ and $C_{s,\mbox{\tiny{max}}}$ are the adsorption and desorption coefficients and the equilibrium foulant concentration, respectively.  All equations are coupled through the boundary conditions defined on the membrane.

The set of Eq.~\eqref{eq:flow}-\eqref{eqn:Cs} allows us to solve for   the dynamical evolution of fouling by coupling the transient equations \eqref{eq:flow}, \eqref{eqn:Cb} and \eqref{eqn:Cs} through the boundary conditions \eqref{BC_membraneV} and \eqref{BC_membraneC}. The primary advantages of the proposed model over more standard methods are the following: (i) all physics is resolved dynamically; (ii) by imposing the condition that $C_{s,\mbox{\tiny{max}}}>C_{\mbox{\tiny{b}}}$, the growth function \eqref{eqn:Cs} guarantees the that $A_f \left(C_{\mbox{\tiny{s}}}-C_{\mbox{\tiny{b}}}\right)>0$, and prevents any unphysical flux increase; (iii) the coupling between  bulk  and foulant concentration, and between the foulant and flux reduction is explicitly modeled and no additional   hypothesis is needed to describe the functional dependence between foulant accumulation and membrane resistance \citep{sanaei2016flow,sanaei2017flow,griffiths2013control}; (iv)  different growth kinetics can be accounted for by appropriately modifying $K_1$ and $K_2$, for instance, soluble foulant (e.g. sodium chloride) has $K_1>0$ and $K_2>0$, while more resilient foulant (e.g. calcium carbonate) has $K_1>0$ and $K_2\approx 0$. 
 
Once the foulant surface concentration $C_{\mbox{\tiny{s}}}(X,Y,H,T)$ is determined, the non-fouled regions $\Gamma_n$ on the RO membrane are identified by locally thresholding $C_{\mbox{\tiny{s}}}$, i.e. 
\begin{equation}
	\Gamma_{n}(T):=\left\{(X(T),Y(T), H) | C_{\mbox{\tiny{s}}}\leq \alpha C_{s,\mbox{\tiny{max}}}\right\}, \quad \mbox{where} \quad \alpha\in[0,1], \,T\in(0,T_{\mbox{\tiny{max}}}).
\end{equation}
which represents the area formed by a set of membrane points $(X,Y,H)$ that satisfy the condition $C_{\mbox{\tiny{s}}}\leq \alpha C_{s,\mbox{\tiny{max}}}$. In this study we set $\alpha =1$. The unit permeate flux $Q_{\mbox{\tiny{m}}}$ [$LT^{-1}$] is
\begin{equation}
	Q_{\mbox{\tiny{m}}}(T)=\dfrac{1}{A_{\mbox{\tiny{m}}}}\int_{\Gamma_{n}(T)} W_{\mbox{\tiny{H}}}(X,Y,T) d\mathcal{A},
	\label{eqn:Flux}
\end{equation}
where $A_{\mbox{\tiny{m}}}$ is the total surface of the membrane.

\subsection{Dimensionless  Formulation}\label{sec:dimensionless}

We start by defining  the Sherwood number, i.e. the ratio between the convective and diffusive mass transport toward the membrane or  the dimensionless permeate flux,
\begin{align}\label{eqn:Sh_def}
\Sh=\dfrac{Q_{m,\infty}B}{D},
\end{align}
where $Q_{m,\infty}$ is the steady-state  permeate flux, and the Bejan number, i.e. the dimensionless pressure head drop along the channel,
\begin{align}\label{eqn:Be_def}
\Be=\dfrac{ B^2 }{\nu^2}\Delta \hat{P},
\end{align}
where $\Delta \hat{P}=\hat{P}_{\mbox{\tiny{in}}}-\hat{P}_{\mbox{\tiny{out}}}$ is the modified pressure  drop along the membrane between the inlet ($X=0$) and the outlet ($X=L$). Furthermore, we define the following dimensionless quantities,
\begin{align}
& \mathbf{u}=\frac{\mathbf U}{U_{\mbox{\tiny{in}}}}, \quad \mathbf{x}=\frac{\mathbf X}{B}, \quad t= \frac{T}{B/U_{\mbox{\tiny{in}}}} \quad P=\frac{\hat P}{\nu^2/B^2}, \quad A_0=\frac{\hat{A}_0}{\nu^2/B^2}, \quad w_{\mbox{\tiny{h}}}=\frac{W_{\mbox{\tiny{H}}}}{U_{\mbox{\tiny{in}}}}, \quad h=\frac{H}{B}
\end{align}
where $\mathbf{u}=(u,v,w)$ and $\mathbf{x}=(x,y,z)$ are the dimensionless velocity field and coordinate axes. We also introduce  the dimensionless numbers
\begin{align}
& \Rey=\frac{U_{\mbox{\tiny{in}}} B}{\nu}, \quad \Pen=\frac{U_{\mbox{\tiny{in}}} B}{D}, \quad \Dc=\frac{k_{\mbox{\tiny{m}}}}{B^2}, \quad \Da_{\mbox{\tiny{I}}}=K_1  \frac{B}{U_{\mbox{\tiny{in}}}}, \quad \Da_{\mbox{\tiny{II}}}=K_2  \frac{B}{U_{\mbox{\tiny{in}}}},
\end{align}
where $Re$, $Pe$, $\Dc$, $\Da_{\mbox{\tiny{I}}}$ and $\Da_{\mbox{\tiny{II}}}$   are the Reynolds, P\'{e}clet, Darcy and Damk\"{o}hler numbers related to the adsorption and desorption reactions, respectively. Then, the transport equations \eqref{eqn:Cb} and \eqref{eqn:Cs} for the bulk and surface concentration,  $C_{\mbox{\tiny{b}}}$ and $C_{\mbox{\tiny{s}}}$, can be cast in dimensionless form
%
\begin{align} 
&\Pen\left( \frac{\partial C_{\mbox{\tiny{b}}}}{\partial t}+\mathbf{u}\cdot \nabla C_{\mbox{\tiny{b}}}\right)-\nabla^2 C_{\mbox{\tiny{b}}}=0 \label{eqn:dim_Cb}
\end{align}
and
\begin{align} 
\frac{\partial C_{\mbox{\tiny{s}}}}{\partial t} &=\Da_{\mbox{\tiny{I}}} (C_{s,\mbox{\tiny{max}}}-C_{\mbox{\tiny{s}}}) C_{\mbox{\tiny{b}}}-\Da_{\mbox{\tiny{II}}}  C_{\mbox{\tiny{s}}}, \label{eqn:dim_Cs}
\end{align}
subject to 
\begin{subequations}\label{eq:dim_BC}
\begin{align} 
 \frac{\partial C_{\mbox{\tiny{b}}}}{\partial z} &=\Pen \, w_{\mbox{\tiny{h}}}  C_{\mbox{\tiny{b}}}, \label{eqn:dim_Cb_BC} \\
 w_{\mbox{\tiny{h}}} &=\frac{\Dc}{ \Rey} \left[ \delta P- A_0\left(C_{\mbox{\tiny{s}}}-C_{\mbox{\tiny{b}}}\right)\right],\label{eqn:dim_W_BC}
\end{align}
\end{subequations}
 on the membrane surface ($z=h$).
 %
%

%
For practical applications, it is often important to identify the relationship between  pressure drop and the permeate flux when the system reaches equilibrium in the long-time limit, i.e.  
\begin{align}
\Sh = \Pi ( \Be, \Rey, \Pen, \Dc, etc. ),
\end{align}
using the formulation above.
In the following section, we will derive an analytical scaling behavior between $Be$ and $Sh$.%

\subsection{Long-time scaling limit}\label{sec:scaling}
At steady-state, Eq.~\eqref{eqn:dim_Cs} reads
\begin{align} \label{eqn:Cs_SS}
0=\Da_{\mbox{\tiny{I}}} (C_{s,\mbox{\tiny{max}}}-C_{\mbox{\tiny{s}}}) C_{\mbox{\tiny{b}}}-\Da_{\mbox{\tiny{II}}}  C_{\mbox{\tiny{s}}},
\end{align}
namely,
\begin{align} \label{eqn:Cs_SS}
C_{\mbox{\tiny{s}}}=\frac{\Da_{\mbox{\tiny{I}}} C_{s,\mbox{\tiny{max}}}C_{\mbox{\tiny{b}}}}{\Da_{\mbox{\tiny{II}}}+\Da_{\mbox{\tiny{I}}}C_{\mbox{\tiny{b}}}}. 
\end{align}
Combining \eqref{eqn:Cs_SS} with the membrane permeate flux equation~\eqref{eqn:dim_W_BC}, we obtain
\begin{equation} \label{eqn:longtimeW}
 w_{\mbox{\tiny{h}}} =\frac{\Dc} {\Rey} \left[ \delta P- A_0\left(\frac{\Da_{\mbox{\tiny{I}}} C_{s,\mbox{\tiny{max}}}C_{\mbox{\tiny{b}}}}{\Da_{\mbox{\tiny{II}}}+\Da_{\mbox{\tiny{I}}}C_{\mbox{\tiny{b}}}}-C_{\mbox{\tiny{b}}}\right)\right].
\end{equation}
Assuming the foulant accumulates much faster than it dissolves, i.e. $\Da_{\mbox{\tiny{I}}}C_{\mbox{\tiny{b}}}/\Da_{\mbox{\tiny{II}}} \ll1$, Eq. \eqref{eqn:longtimeW} can be simplified as follows
\begin{equation}\label{bc:intermediate2}
 w_{\mbox{\tiny{h}}}= \frac{\Dc} {\Rey} \left[ \delta P- A_0\left(C_{s,\mbox{\tiny{max}}}-C_{\mbox{\tiny{b}}}\right)\right].
\end{equation}
Also, under the assumption that at steady state  $P(z=h^-)\sim P_{\mbox{\tiny{in}}}$, while accounting for Eq.~\eqref{eqn:Be_def},  Eq.~\eqref{eqn:dp} can be written as
\begin{equation}\label{inter2} 
\delta P=\Be +P_{\mbox{\tiny{out}}},
\end{equation}
where $P_{\mbox{\tiny{in}}}$ and $P_{\mbox{\tiny{out}}}$ are the dimensionless inlet and outlet pressures.
Inserting \eqref{inter2}  into \eqref{bc:intermediate2},  we obtain
\begin{equation}\label{bc:intermediate3}
 w_{\mbox{\tiny{h}}}= \frac{\Dc} {\Rey} \left[\Be +P_{\mbox{\tiny{out}}}- A_0\left(C_{s,\mbox{\tiny{max}}}-C_{\mbox{\tiny{b}}}\right)\right].
\end{equation}
Under the hypothesis that $C_{\mbox{\tiny{b}}}\approx C_{\mbox{\tiny{b}}}(Z)$, i.e. the variation of the bulk concentration $C_{\mbox{\tiny{b}}}$ with  $x$  and $y$ is negligible, while accounting for \eqref{bc:intermediate3}, the boundary condition~\eqref{eqn:dim_Cb_BC}  can be written as 
%
%
%
\begin{equation}\label{eqn:šCb_ode}
C_{\mbox{\tiny{b}}}' =\Pi_{\mbox{\tiny{I}}} C_{\mbox{\tiny{b}}}^2+ \Pi_{\mbox{\tiny{II}}}   C_{\mbox{\tiny{b}}} ,
\end{equation}
where
\begin{subequations}
\begin{align}
\Pi_{\mbox{\tiny{I}}} & = \Dc \, Sc \, A_0,\\
\Pi_{\mbox{\tiny{II}}} & =\Dc \, Sc \left( \Be-\Be^{\star}\right),\label{Pi_2}
\end{align}
\end{subequations}
and
\begin{align} 
\Be^{\star}&=A_0 C_{s,\mbox{\tiny{max}}} -P_{\mbox{\tiny{out}}},
\end{align}
since $Sc=Pe/Re=\nu/D$.
%
Eq. \eqref{eqn:šCb_ode} is a homogeneous non-linear ordinary differential equation for $C_{\mbox{\tiny{b}}}$, which can be solved by using the following substitution:
\begin{equation}
\gamma=\frac{1}{C_{\mbox{\tiny{b}}}}.
\end{equation}
The transformed equation gives us a non-homogeneous equation for $\gamma$ 
\begin{equation}
-\gamma' =  \Pi_{\mbox{\tiny{I}}} + \Pi_{\mbox{\tiny{II}}}\gamma,
\end{equation}
whose solution, $\gamma=\gamma_h+\gamma_p$, is given by the general and particular solutions, $\gamma_h$ and $\gamma_p$, which satisfy
\begin{equation}
-\gamma_h' =\gamma_h \Pi_{\mbox{\tiny{II}}}
\end{equation}
and 
 \begin{equation}
-\gamma_p' =\Pi_{\mbox{\tiny{I}}} + \gamma_p \Pi_{\mbox{\tiny{II}}} ,
\end{equation}
respectively. The solution is
\begin{equation}
\gamma=\gamma_h+\gamma_p=C_1 \exp\left(-\Pi_{\mbox{\tiny{II}}}z\right)-\frac{\Pi_{\mbox{\tiny{I}}}}{\Pi_{\mbox{\tiny{II}}}},
\end{equation}
where $C_1$ is  determined by imposing the boundary condition
\begin{equation}
C_{\mbox{\tiny{b}}}=1, \quad \mbox{when} \quad z=0,
\end{equation}
i.e. the channel bottom  reaches saturation at  steady-state. The  solution reads
\begin{equation}\label{eq:c_bfinal}
C_{\mbox{\tiny{b}}}=\frac{\Pi_{\mbox{\tiny{II}}}}{\left( \Pi_{\mbox{\tiny{I}}} + \Pi_{\mbox{\tiny{II}}}\right)+\exp\left(-\Pi_{\mbox{\tiny{II}}}h\right) -\Pi_{\mbox{\tiny{I}}}}.
\end{equation}
\begin{figure}
  \centerline{\includegraphics[width=\textwidth]{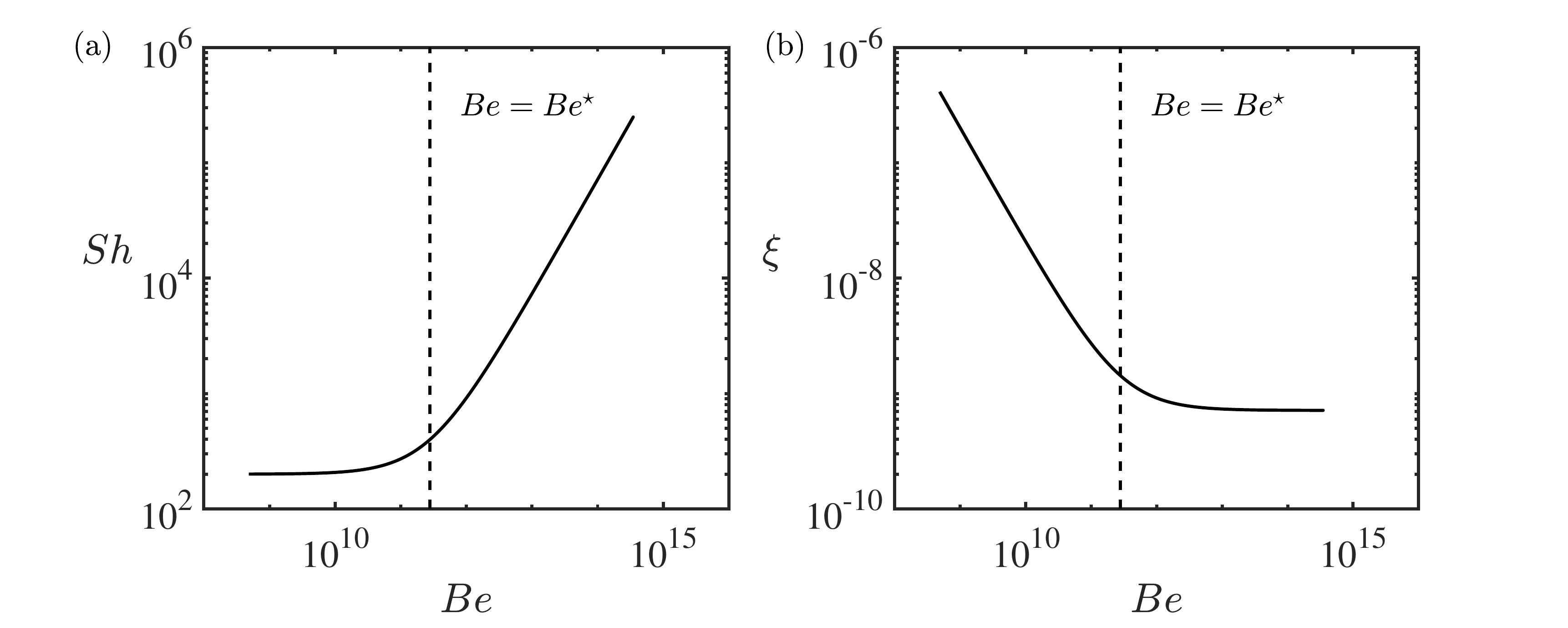}}
  \caption{(a) Analytical solution of Sherwood number $\Sh$ as a function of Bejan number $\Be$ for parameters values listed in Table~\ref{table:sim_parameters}; (b) Analytical solution of the efficiency $\xi$ as a function of Bejan number.}
\label{fig:Sh_vs_Be_Analytical}
\end{figure}
Additionally, we assume that at steady state $w_{\mbox{\tiny{h}}}=Q_{m,\infty}/U_{\mbox{\tiny{in}}}$, i.e.
\begin{equation}\label{eq:wh_dim}
w_{\mbox{\tiny{h}}}=\frac{1}{\Pen}\Sh.
\end{equation}
Combining \eqref{eq:wh_dim} with \eqref{bc:intermediate3}, we obtain
\begin{equation}
\Sh = \Pi_{\mbox{\tiny{I}}} C_{\mbox{\tiny{b}}} + \Pi_{\mbox{\tiny{II}}}.
\end{equation}
Evaluating $C_{\mbox{\tiny{b}}}$ at the membrane surface, $z=h$, while accounting  \eqref{eq:c_bfinal}, leads to
\begin{equation}\label{sol}
\Sh = \Pi_{\mbox{\tiny{II}}} - \frac{\Pi_{\mbox{\tiny{II}}}}{1- \left( \frac{\Pi_{\mbox{\tiny{II}}}}{\Pi_{\mbox{\tiny{I}}}}+1\right)-\exp\left(-\Pi_{\mbox{\tiny{II}}}h\right)}. 
\end{equation}
Substituting \eqref{Pi_2}, we obtain
\begin{equation} \label{eqn:longtimeW_Dimless_2}
\Sh = \Dc \, Sc(\Be-\Be^\star)\left\{1-\frac{1}{1-[1+\frac{1}{\Pi_{\mbox{\tiny{I}}}}(\Be-\Be^\star)]\exp\left[ -h \Dc \, Sc(\Be-\Be^\star)\right]}\right\},
\end{equation}
which provides the relationship between the dimensionless permeate flux, $\Sh$, and the dimensionless pressure drop, $\Be$. It is worth emphasizing that, at steady state, $\Sh$ can be written as a function of only Bejam number $\Be$,  Darcy number, $\Dc$, i.e. the dimensionless permeability of the membrane, and  Schmidt number $Sc$ (as well as the geometric parameter $h$), while it is independent of $Re$, $Pe$ and $Da$. However, since $\Be$ is certainly function of $Re$ (as $\Sh$ is), \eqref{eqn:longtimeW_Dimless_2} implies that both $\Sh$ and $\Be$ exhibit the same scaling behavior in terms of Reynolds number.

We now look at the asymptotic behavior of \eqref{eqn:longtimeW_Dimless_2} for $\Be\rightarrow 0$ and  $\Be\rightarrow \infty$, while keeping all the other dimensionless parameters constant, and obtain
%
%
\begin{subequations}\label{scaling}
\begin{align}
&\lim_{\Be \to 0} \Sh \sim \mbox{const},\label{Be-0}\\
&\lim_{\Be \to \infty} \Sh \sim  \Be,\label{sh1}
\end{align}
\end{subequations}
with the transition between the two scaling behaviors occurring at 
\begin{equation}\label{sh2}
\Be\sim\Be^\star.
\end{equation}
%
Since  $\Be$ can be associated with the energy input for filtration operations and $\Sh$ is the quantity to maximize, we define the \emph{overall filtration performance index}, $\xi$, as:
\begin{align} \label{eqn:efficiency}
\xi=\dfrac{\Sh}{\Be},
\end{align}
where the higher the value of $\xi$, the higher the membrane efficiency both in terms of permeate flux and energy input.  Combining \eqref{eqn:efficiency} with \eqref{Be-0} and \eqref{sh1}, one obtains
\begin{align}\label{xi1}
\xi\sim \dfrac{1}{\Be}, \quad \Be\rightarrow 0,
\end{align}
and
\begin{align}\label{xi2}
\xi\sim \mbox{const} \quad \Be\rightarrow \infty,
\end{align}
respectively.  Figures~\ref{fig:Sh_vs_Be_Analytical}(a) and (b) show the relationship between the Sherwood  and the Bejan numbers as defined by Eq.~\eqref{eqn:longtimeW_Dimless_2} for the set of parameters listed in  Table~\ref{table:sim_parameters}, and the membrane performance index $\xi$, respectively. The dashed line is the transition $\Be=\Be^\star$ between the two scaling behaviors. The scalings \eqref{scaling}-\eqref{xi2} suggest that  an increase in the inlet velocity (or, equivalently, pressure drop) leads to an increase in the permeate flux after certain threshold ($\Be^\star$) is overcome. However, the overall system performance drops as a result  and reaches a plateau when $\Be \to \infty$: at high flow rates, the increased energy requirement to sustain a given pressure drop outweighs any benefits due to reduced fouling. This suggests (and will be confirmed in the following) that under these conditions, surface topology modifications may  better impact membrane performance  than morphological changes (e.g. the adoption of spacers). Instead,  when $\Be \to 0$, $\Sh$ does not depend strongly on the pressure drop. 
%

While the scaling behavior \eqref{xi1} and \eqref{xi2} is obtained for the benchmark case of a rectangular membrane, in the following we move forward by, first, validating the proposed model equations against experimental results on fouled rectangular membranes (\S\ref{sec:numerical}), and then generalize the approach to membranes with  complex morphological and topological features (i.e., additional length scales) (\S\ref{sec:synthetic}).

\section{Numerical model validation}\label{sec:numerical}
In the following, we proceed by validating (1) the model  \eqref{eq:flow}-\eqref{eqn:Cs}  against experimental data and (2) the scaling relationships between the quantities of interest.

\begin{figure}
  \centerline{\includegraphics[width=\textwidth]{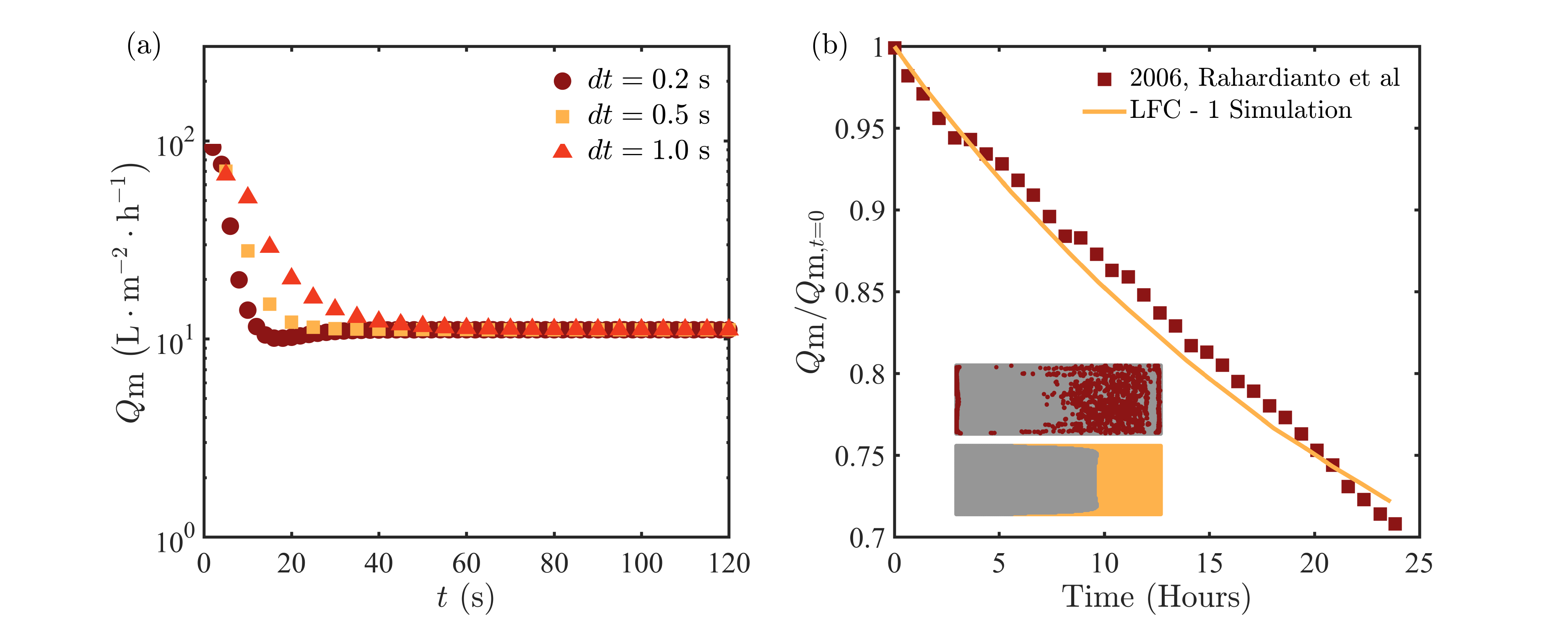}}
  \caption{(a) Simulated permeate flux for different time steps. Results show that the steady-state solution is reached around $T\approx 60$ s; (b) Comparison between dynamic measurements of permeate flux collected by  \cite{rahardianto2006diagnostic} (symbols) and simulated permeate flux decrease (solid line).  The inset shows a comparison between the digitalized experimental steady-state fouling pattern (top) and the simulated one (bottom).}
\label{fig:validation}
\end{figure}

\subsection{Implementation and Convergence Study}\label{sec:convergence}

We implement \eqref{eq:flow}-\eqref{eqn:Cs} in the Finite-Volume OpenFOAM\textsuperscript{\textregistered} framework, by developing the customized solver SUMs (Stanford University Membrane solver). The solver is explicit in time  and second order in space. A convergence study is performed using a straight channel of dimensions $H\times L\times B$=$2\times70\times5$ mm$^3$. The permeate flux across the membrane $Q_m(T)$ (Liter m$^{-2}$hour$^{-1}$)  is  calculated using three different time steps $dT=\{0.2,0.5,1.0\}$ s, and simulated for $T_{\mbox{\tiny{max}}}=120\,\mbox{s}$. Figure~\ref{fig:validation} shows the calculated $Q_m(T)$ for the three different scenarios. The three transient simulations converge to the same steady state, independently of the time step. The steady-state solution is achieved at $T\approx60\,\mbox{s}$. For  simulations involving non-rectangular geometries, we  run the simulations for $400 \mbox{ s}$ to ensure  steady-state is reached.  In the next section, we proceed with validating the code against fouling experiments.

\begin{table}
\centering
\begin{tabular}{c c c c c c c c c c}
         \multirow{ 2}{*}{Parameters } & $U_{\mbox{\tiny{in}}}$ & $\Rey$ & $\Rey^{\star}$ & $k_{\mbox{\tiny{m}}}$   & $K_{1} $& $K_{2}$ & $C_{\mbox{\tiny{s,max}}}$ & $\hat{P}_{\mbox{\tiny{out}}}$ & $\hat{A}_0$\\	
					                            & $[\mbox{m}/\mbox{s}]$ & $[-]$& $[\times 10^{3}]$ & $[\mbox{mD}]$ & $[1/\mbox{s}]$& $[1/\mbox{s}]$ & $[-]$ & $[\mbox{m}^2/\mbox{s}^2]$ & $[\mbox{m}^2/\mbox{s}^2] $\\	
			     \hline
				 Rahardianto \emph{et al.} & $0.15$ & $500$ & - & $6.95$ & - & - & - & 1000 & - \\
				 \hline
				 LFC - 1 & $0.15$ & $500$ & - & $6.95$ & $1\times10^{-5}$ & $1\times10^{-7}$ & 2 & 1000 & 100\\
				 \hline
         SIM - 1 & $0.05$ & 250& $0.25 - 3$  & \multirow{ 6}{*}{$7.00$}  & \multirow{6}{*}{0.1} & \multirow{6}{*}{0.001} & \multirow{ 6}{*}{2} & \multirow{ 6}{*}{1000} & \multirow{ 6}{*}{100}\\
				 SIM - 2 & $0.075$  & 375 & $0.38 - 5$  & &  & & &\\
				 SIM - 3 & $0.1$  & 500 & $0.5 - 7$ & &  & & &\\
				 SIM - 4 & $0.15$  & 750 & $0.75 - 10$ & & & & &\\
         SIM - 5 & $0.175$  & 875 & $0.87 - 13$ & & & & &\\
				 SIM - 6 & $0.2$ & 1000 & $10 - 15$ &  & & & &\\
	       \hline
\end{tabular}
	\caption{Parameters of all numerical simulations. First row: parameters from  the experimental study by Rahardianto \emph{et al.}; Other rows: parameters of the synthetic examples of Section~\ref{sec:synthetic} with Darcy number $Dc=1.967 \times 10^{-10}$ and Schmidt number $Sc=500$.}
  \label{table:sim_parameters}
\end{table}

\subsection{Experimental Validation}\label{sec:validation}
To validate the model proposed in Section~\ref{sec:model} and the developed solver, we compare numerical simulations of permeate flux decrease over time and fouling development with membrane fouling experiments conducted by \cite{rahardianto2006diagnostic}. In \cite{rahardianto2006diagnostic}'s study,  fouling experiments are performed using a Low Fouling Composite (LFC) Membrane with crystallized gypsum as the foulant. Both steady-state and transient measurements of permeate flux, as well as final fouling patterns, are provided. 


The simulation parameters are set to match the experimental set-up and operating conditions. Table~\ref{fig:validation} lists all the experimental parameters used in the simulation. Since measurements of the membrane permeability  are not provided, $k_{\mbox{\tiny{m}}}$ is estimated from  pressure and flux measured during a clean water experiment  through \eqref{BC_membraneV}, where $C_{\mbox{\tiny{b}}}$ and $C_{\mbox{\tiny{s}}}$ are set to zero. The  coefficient $A_0$ is fitted  to match the initial permeate flux at $T=0$, $Q_{m}(T=0)$; it is then  kept constant throughout the simulation.

The coefficients $K_1$, $K_2$ and $C_{s,\mbox{\tiny{max}}}$ in \eqref{eqn:Cs}, not provided by \cite{rahardianto2006diagnostic}, are estimated as follows: following the experimental observations by \cite{xie2014hydrodynamics} where  foulant accumulated on the membrane is approximately $C_{s,\mbox{\tiny{max}}}=1.44 C_0$ (i.e.  $2-4$ times the bulk concentration) for $C_0=0.4-0.6\,\mbox{M}$, we set the equilibrium foulant concentration  to $C_{s,\mbox{\tiny{max}}}=2C_0$.
The membrane adsorption/desorption rate $\theta$, i.e. the ratio between $K_2$ and $K_1$, $\theta:=K_2/K_1$ varies from $0.1$ to $0.001$ \citep{jones2000protein}. In our study, we set  $\theta=0.01$, with  $K_1=1\times10^{-5}$ and  $K_2=1\times10^{-7}$. 

Figure~\ref{fig:validation}(b) shows the  comparison between the numerical predictions and the experimental measurements of the  permeate flux decline as a function of time. The measured and predicted  foulant spatial patterns are shown in the inset of Figure~\ref{fig:validation}(b). The comparison demonstrates that the system \eqref{eq:flow}-\eqref{eqn:Cs}  can correctly capture both unsteady effects as well as the spatio-temporal evolution of foulant accumulation.

In the following, we perform a series of unsteady fully 3D numerical studies to assess and elucidate the impact that different modifications of the filtration system have on fouling. We classify them into two broad classes: (i) \emph{morphological} changes  entail modifications of the design of the flow channel (i.e. the spacer morphology) and have  characteristic length scales in the order of mm; instead, (ii) \emph{topological} alterations introduce micro-scale  patterns/features on the membrane surface and have characteristic length scales ($\mu$m or sub-$\mu$m) that are much smaller than the channel dimensions.
\section{Impact of membrane morphology and topology on fouling control}\label{sec:synthetic}


\begin{figure}
  \centerline{\includegraphics[width=\textwidth]{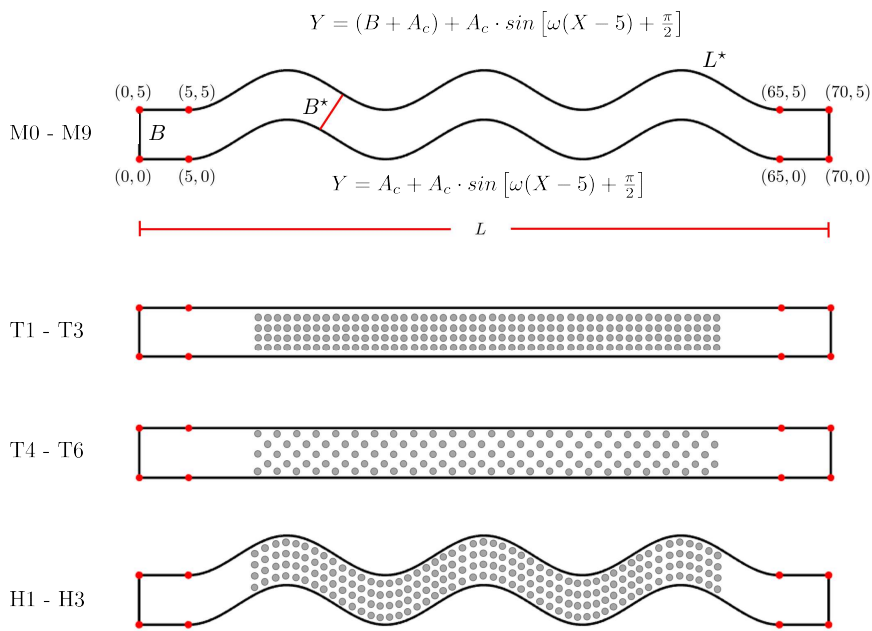}}
  \caption{Schematic of the channel geometries investigated. Designs M0-M9 involve morphological changes (i.e. various sinusoidal shapes); Designs T1-T6 involve topological changes where the membrane is patterned with pillars of different heights and arrangements;  H1-H3 are hybrid designs which combine both morphological and topological alteration of the membrane.}
\label{fig:Geometry}
\end{figure}

\begin{table}
\centering
\setlength{\tabcolsep}{15pt}
\renewcommand{\arraystretch}{1.2}
\begin{tabular}{r c c c c c}
         \hline
         No. & $A_c[\mbox{mm}]$ & $\omega$ & $B^{\star} [\mbox{mm}]$ & $L^{\star}[\mbox{mm}]$ & $H^{\star}[\mbox{mm}]$\\		
				 \hline
         M0 & $0$ & $0$ & $5.00$ & 70.00 & 0\\
				 M1 & $2$ & $\pi/30$ & $4.89$ & 70.65 & 0\\
				 M2 & $4$ & $\pi/30$ & $4.61$ & 72.55 & 0\\
				 M3 & $6$ & $\pi/30$ & $4.24$ & 75.55 & 0\\
				 M4 & $2$ & $\pi/10$ & $4.26$ & 75.55 & 0\\
				 M5 & $4$ & $\pi/10$ & $3.16$ & 89.25 & 0\\
				 M6 & $6$ & $\pi/10$ & $2.36$ & 107.12 & 0\\
				 M7 & $2$ & $\pi/5$ & $3.29$  & 89.25 & 0\\
				 M8 & $4$ & $\pi/5$ & $1.93$  & 127.12 & 0\\
				 M9 & $6$ & $\pi/5$ & $1.31$  & 170.20 & 0\\
				 T1 & $0$ & $0$ & $5.00$ & 70.00 & 0.5\\
				 T2 & $0$ & $0$ & $5.00$ & 70.00 & 1.0\\
				 T3 & $0$ & $0$ & $5.00$ & 70.00 & 1.5\\
				 T4 & $0$ & $0$ & $5.00$ & 70.00 & 0.5\\
				 T5 & $0$ & $0$ & $5.00$ & 70.00 & 1.0\\
				 T6 & $0$ & $0$ & $5.00$ & 70.00 & 1.5\\
				 H1 & $2$ & $\pi/10$ & $4.26$ & 75.55 & 0.5\\
				 H2 & $2$ & $\pi/10$ & $4.26$ & 75.55 & 1.0\\
				 H3 & $2$ & $\pi/10$ & $4.26$ & 75.55 & 1.5\\
				 \hline
\end{tabular}
	\caption{Geometry of channel spacers}
  \label{table:geometry}
\end{table}
\subsection{Numerical simulations}\label{sec:numerical_modified}
We study 18 membrane designs which include 9 purely morphological (M1 to M9), 6 purely topological (T1 to T6) and 3 hybrid designs (H7 to H9)  which include both topological and morphological modifications (see Figure~\ref{fig:Geometry}). The fully  three-dimensional  domains  contain 500,000$-$1,000,000 finite-volume cells.  A  smooth straight channel design (M0) is modeled as the reference case. The morphology of choice in this study is sinusoidal channels of different periods and amplitudes \citep{xie2014hydrodynamics}. For the designs M1-M9, the membrane shape is defined by the following bottom and top boundaries in the ($X,Y$)-plane,
\begin{subequations}
\begin{align}
&Y=(A_c+B)+A_c\sin[\omega(X-B)+\pi/2],\\
&Y=A_c+A_c\sin[\omega(X-B)+\pi/2],
\end{align}
\end{subequations}
where $A_c$ and $\omega$  are the amplitude and period of the sinusoidal wave, respectively, and $B$ is the membrane width. The designs T1-T6 are characterized by micropatterns composed by cylindrical posts of different heights and arrangements: T1, T2 and T3 have  square (aligned) patterns with micropillars of different heights, while T4, T5 and T6 designs are characterized by  staggered (hexagonal)  patterns with three different pillar heights.  The hybrid designs H7, H8 and H9 are a combination of morphological and topological changes  with three different pattern heights. Details of all the geometries are given in Figure~\ref{fig:Geometry} and Table \ref{table:geometry}.  For each geometry, we investigate the membrane response to fouling for  six different inlet velocities ($U_{\mbox{\tiny{in}}}$) with Reynolds number, 
\begin{align}
\Rey=\dfrac{U_{\mbox{\tiny{in}}}B}{\nu},
\end{align}
ranging from 250 to 1000, as listed in Table~\ref{table:sim_parameters}, for a total of 114 simulations. 

\begin{figure}
  \centerline{\includegraphics[width=0.9\textwidth]{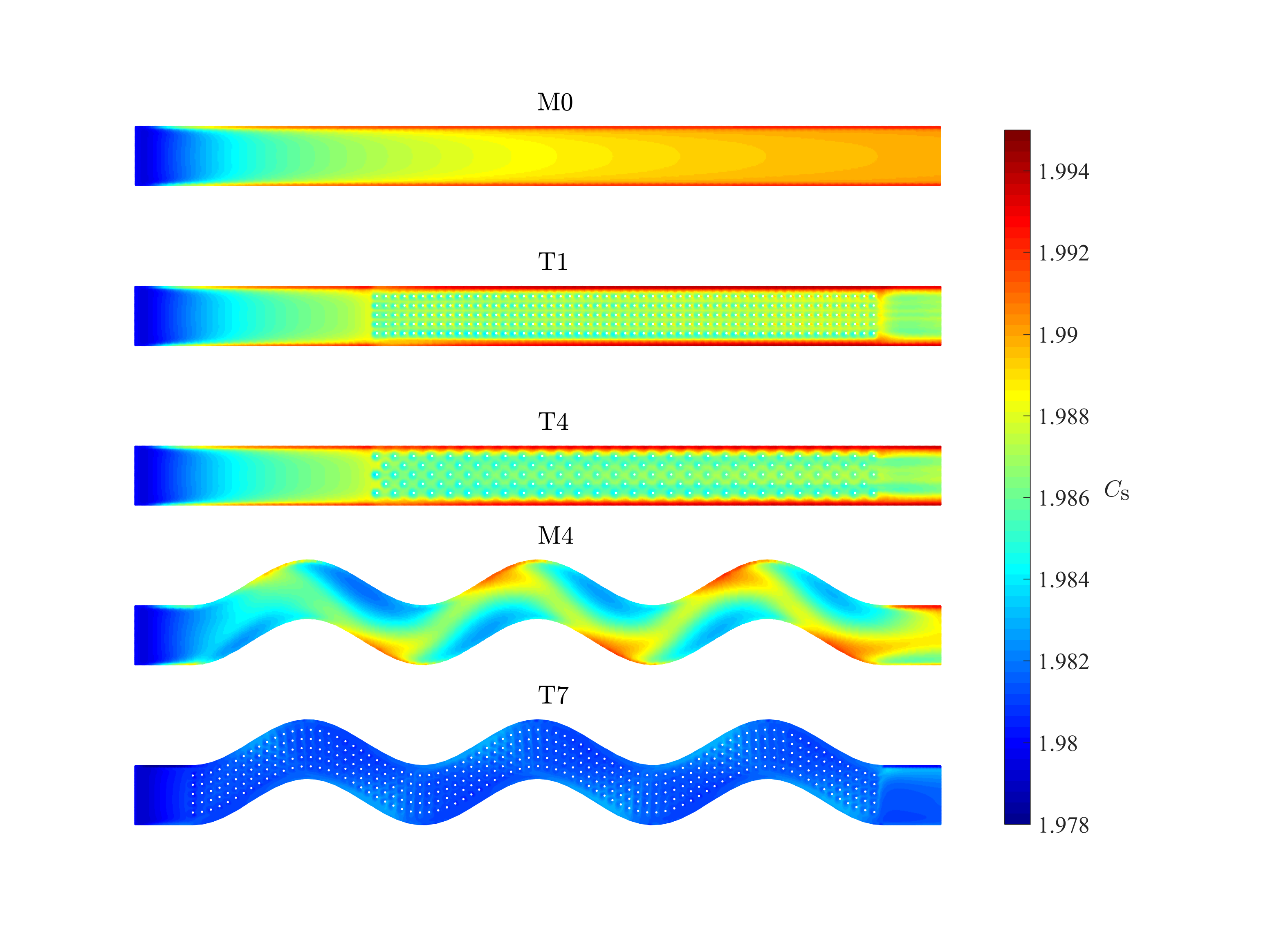}}
  \caption{Steady-state foulant concentration field $C_s$  for M0, T1, T4, M4 and H1 designs and  inlet velocity $U_{\mbox{\tiny{in}}}=0.075\,\mbox{m}/\mbox{s}$.}
\label{fig:Set02_2D}
\end{figure}

\begin{figure}
  \centerline{\includegraphics[width=0.9\textwidth]{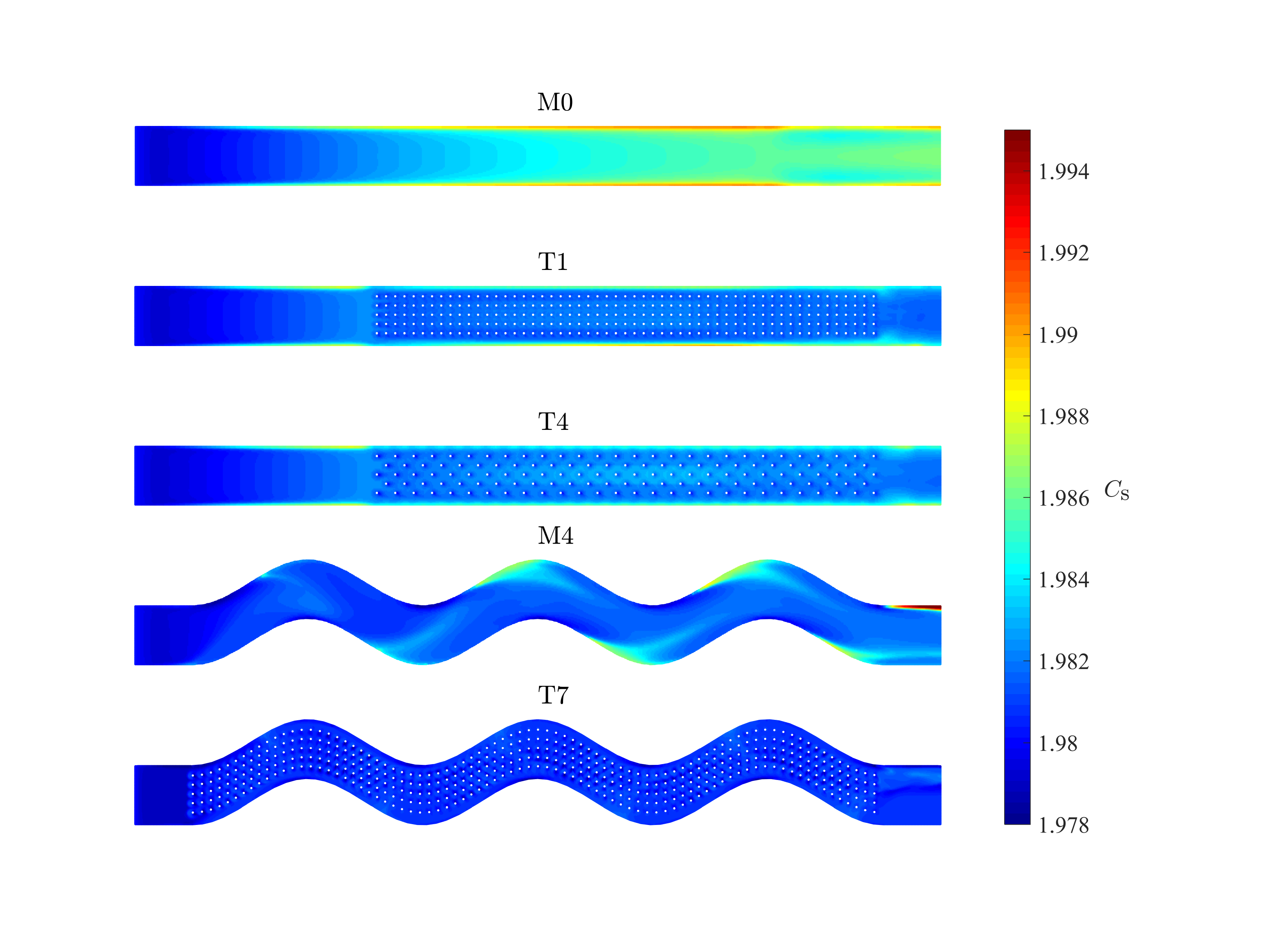}}
  \caption{Steady-state foulant concentration field $C_s$ for M0, T1, T4, M4 and H1 and inlet velocity $U_{\mbox{\tiny{in}}}=0.2\,\mbox{m}/\mbox{s}$.}
\label{fig:Set06_2D}
\end{figure}

Example results are shown in Figures~\ref{fig:Set02_2D} and~\ref{fig:Set06_2D} which provide the spatial distribution of the foulant for membrane types M0, T1, T4, M4 and H7 and two different inlet velocities. Figure~\ref{fig:Set02_2D}  demonstrates how morphological or topological changes of the membrane can significantly   impact foulant distribution. Additionally, a comparison between Figure~\ref{fig:Set02_2D} and Figure~\ref{fig:Set06_2D} suggests that higher inlet velocities significantly decrease foulant accumulation. This is expected since higher inlet (and local) velocities are associated with increased shear stress on the membrane, and reduced foulant accumulation.


\begin{figure}
  \centerline{\includegraphics[width=\textwidth]{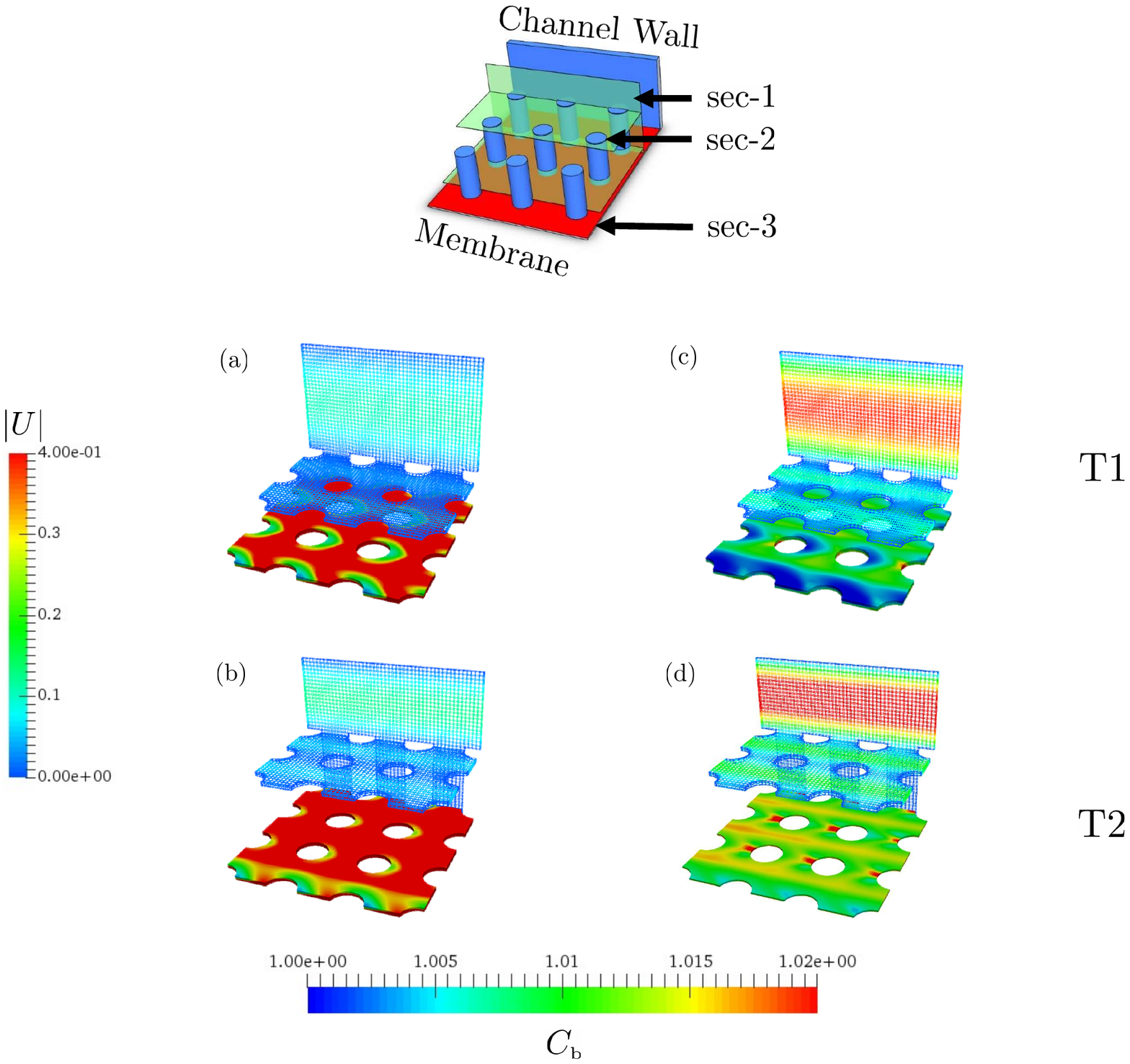}}
  \caption{Velocity (Sections 1 and 2) and  concentration (Section 3) distribution in three sections of channels T1 and T2,  for two different inlet velocities ($U_{in}=0.075\,\mbox{m}/\mbox{s}$ in subplots (a) and (b), and $U_{in}=0.2\,\mbox{m}/\mbox{s}$ in subplots (c) and (d)). The sections are extracted as follows: Section 1 and Section 2 are a vertical section  (parallel to the $X-Z$ plane) and a horizontal section (parallel to the $X-Y$ plane and in proximity of the pattern's top) and show the velocity distribution; Section 3, horizontal and in proximity of the membrane, shows $C_b$.}
\label{fig:Layers}
\end{figure}
%
%
Figure \ref{fig:Layers} shows the velocity (Sections 1 and 2) and  concentration (Section 3) distribution in three sections of channels T1 and T2,  for two different inlet velocities ($U_{in}=0.075\,\mbox{m}/\mbox{s}$ in subplots (a) and (b), and $U_{in}=0.2\,\mbox{m}/\mbox{s}$ in subplots (c) and (d)). The sections are extracted as follows: Section 1 and Section 2 are a vertical section  (parallel to the $X-Z$ plane) and a horizontal section (parallel to the $X-Y$ plane and in proximity of the pattern's top) and show the velocity distribution; Section 3, horizontal and in proximity of the membrane, shows $C_b$. For lower velocities (Figure \ref{fig:Layers} (a) and (b)), T1 and T2 both exhibit strong concentration polarization near the membrane surface with small velocity at the interface: with lower flow rates through the pattern, shear stress on the membrane decreases and foulant accumulation is promoted. Instead,  for higher inlet velocities (Figure \ref{fig:Layers} (c) and (d)), the bulk concentration is smaller near the membrane surface for both cases. However, it is worth noticing that taller pillars, as in the T2 membrane, locally decelerate the flow and reduce antifouling efficiency, compared to their shorter counterparts (T1) where  advective mixing near the membrane significantly reduces C.P. while providing lower flow resistance (and, consequently, pressure drop).
Corresponding fouling patterns ($C_{\mbox{\tiny{s}}}$ distribution) are shown in Figure \ref{fig:Set02_2D} and \ref{fig:Set06_2D}. 

\begin{figure}
  \centerline{\includegraphics[width=0.9\textwidth]{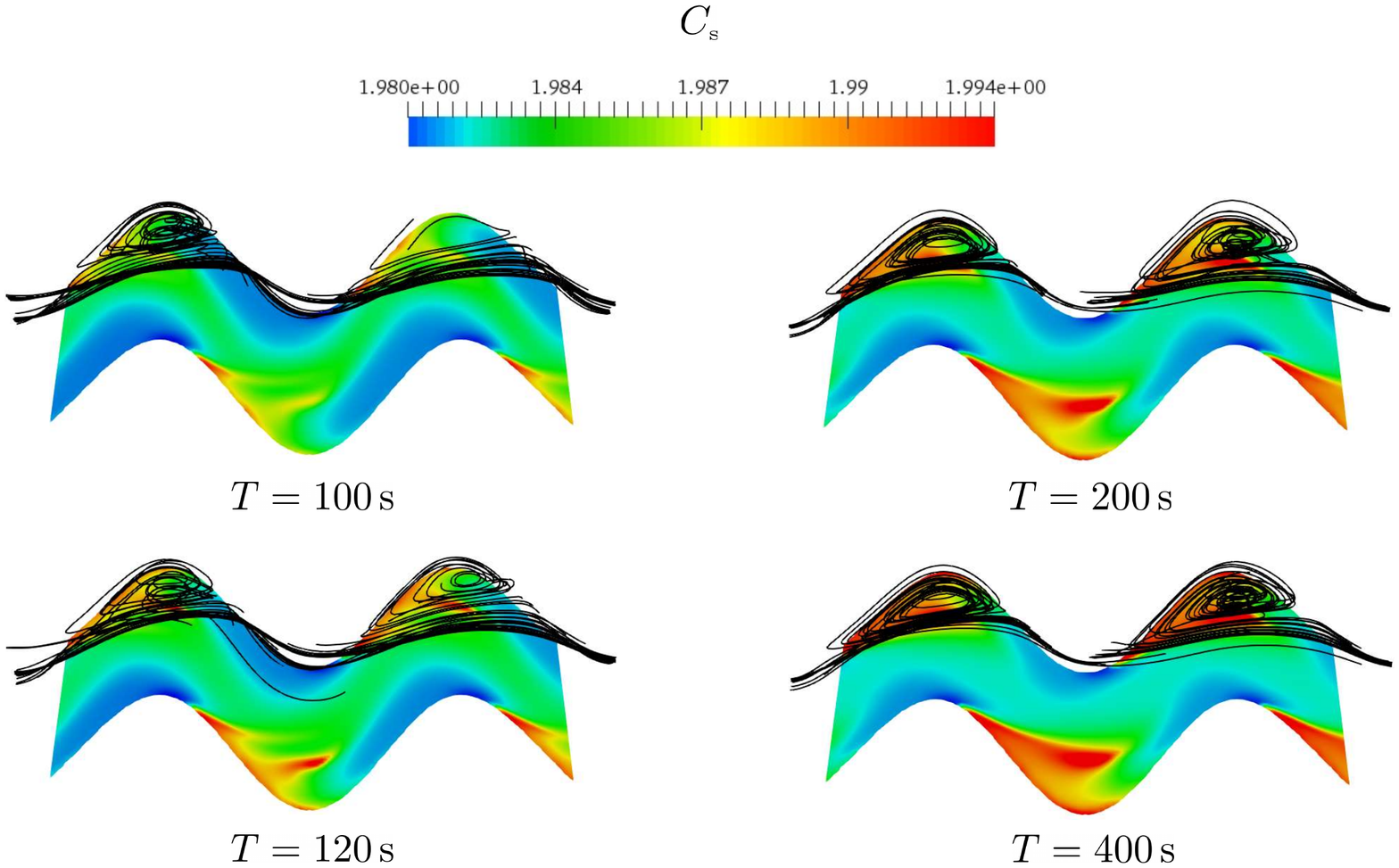}}
  \caption{Foulant distribution $C_s$ (color field) overlaid with streamlines (solid lines) in a portion of channel M7, at four different instances in time ($t=100, 120, 200, 400$ s) and for an inlet velocity $U_{in}=0.075$ m/s.}
\label{fig:Streamlines}
\end{figure}

Additionally, unsteady simulations allow one to  explore where the fouling initiates and how the foulant grows. We extract one section of the flow channel M7 for the simulation SIM-2, and plot the foulant  concentration on the membrane surface $C_s$ together with the streamlines in the channel at different instances in time, see Figure~\ref{fig:Streamlines}. 
Figure \ref{fig:Streamlines} shows that when the flow field is developing, vortices form in the crests and troughs of the sinusoidal channel. At $t\approx 120 \mbox{ s}$, foulant starts to accumulate in the channel, by first nucleating  at the center of the vortex. As time evolves, the foulant accumulation grows following  a spatial pattern similar to that of the vortex.

\subsection{Scaling variables for rough and wiggly membranes}\label{sec:scaling_variable}
The analysis of Section~\ref{sec:dimensionless} provides a useful, although incomplete, framework to study membrane performance: in presence of topological or morphological alterations of the membrane, additional length scales are introduced in the problem, which are not taken into account in the previous analysis. Furthermore, since the input velocity and the length scales associated with the membrane alteration  are the primary decision/design variables, an explicit dependence of the filtration performance index $\xi$ on Reynolds number is desirable.

In order to quantitatively compare the impact that morphological and topological changes have on fouling, we define the following dimensionless length scales,
\begin{align}
\eta=B^\star/B, \quad \mbox{and} \quad \zeta=(H-H^\star)/H
\end{align} 
where $0<\eta\leq 1$ and $0 \leq\xi\leq 1$,  where $B^\star$ is the closest distance between the channel side walls and $H^\star$ is the height of the pattern in the $Z$-direction.  For a straight channel $\eta=1$, and for a topologically unaltered membrane $\zeta=1$, i.e. $\eta$ and $\zeta$ provide a measurement of the `waviness' of the channel and of the `roughness' of the membrane surface, respectively.
Specifically,  $\eta$ and $\zeta$  represent the thinnest channel neck versus the largest width that fluid can experience in $(XY)-$ and $(YZ)-$planes, respectively. Furthermore, we introduce a modified Reynolds number $\Rey^\star$, which accounts for topological and morphological features,
\begin{equation}\label{eq:Re_star}
	\Rey^\star= \left(\frac{1}{\eta^2 \zeta}\right) \Rey.
\end{equation}
In the following, we will show that the modified Reynolds number $\Rey^\star$ allows one to 	quantitatively compare the performance of membranes with different morphological and topological features under a unified framework. In fact, while $Be$ and $Sh$ numbers represent direct estimators of membranes performance, $\Rey^\star$ is the primary decision variable as it concurrently prescribes inlet velocity/volumetric flux and membrane geometry.  

\begin{figure}
  \centerline{\includegraphics[width=0.9\textwidth]{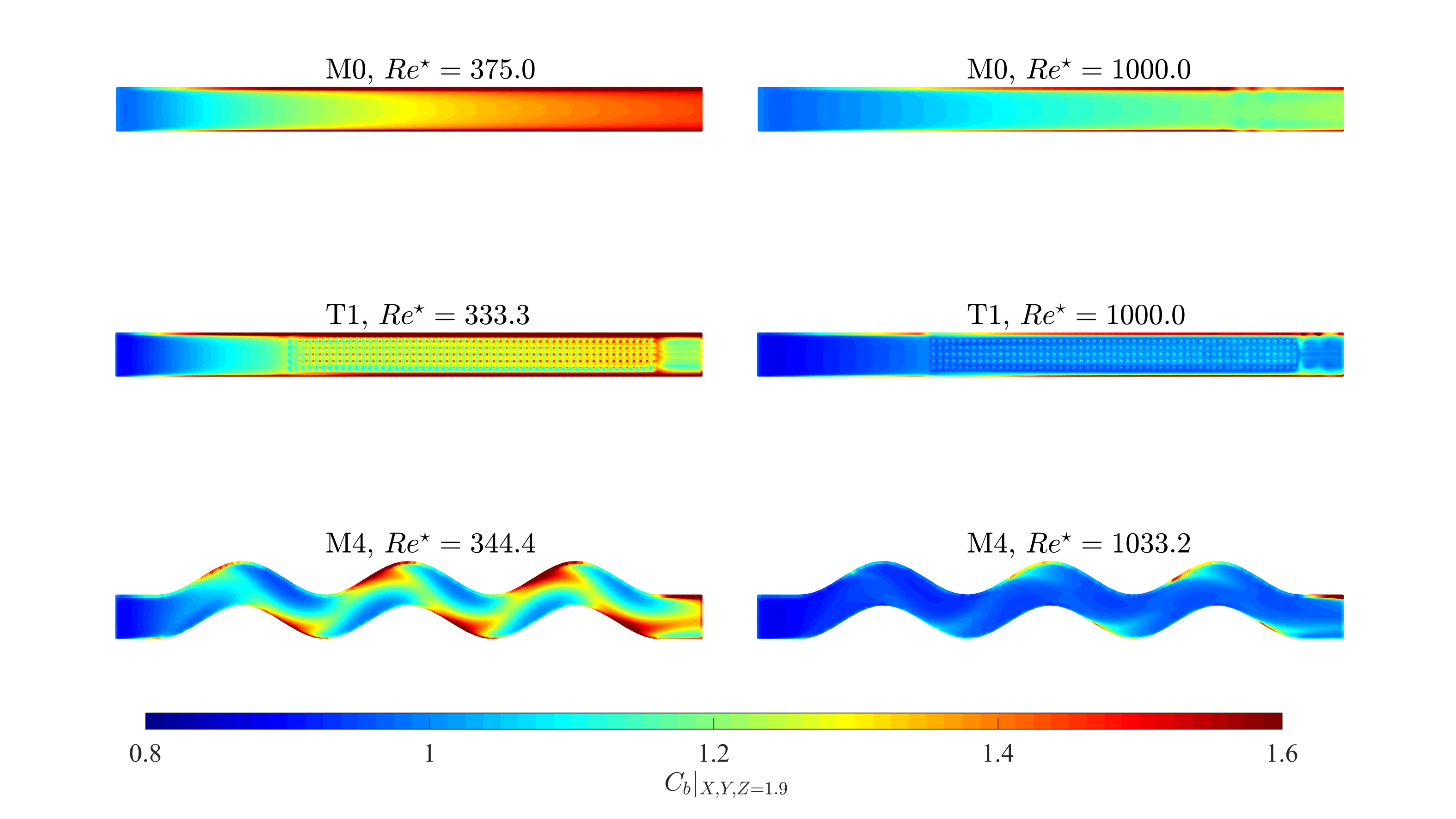}}
  \caption{Bulk concentration  distribution ($C_{\mbox{\tiny{b}}}$) near the membrane surface ($Z=1.9\, \mbox{mm}$) for two different $\Rey^\star$,  $\Rey^\star\approx 300$ (left) and $\Rey^\star\approx 1000$ (right), and three designs, M0 (top), T1 (center) and M4 (bottom).}
\label{fig:Cb_Re_star_2D}
\end{figure}
\subsection{Scaling laws}\label{sec:scaling_laws}

Since Eq.~\eqref{eqn:longtimeW_Dimless_2} suggests that both $Be$ and $Sh$ have the same scaling in terms of Reynolds,  in Figure~\ref{fig:Be_Sh_xi_vs_Omega} we  plot $\Sh$ and $\Be$ as a function of  $\Rey^\star$  for all 114 simulations. In the insets of Figure~\ref{fig:Be_Sh_xi_vs_Omega}, we provide a plot of $\Sh$ and $\Be$ in terms of $\Rey$ for comparison.  Figure~\ref{fig:Be_Sh_xi_vs_Omega}, where all the data points collapse onto one scaling curve,  suggests that $\Rey^\star$ is an appropriate scaling variable, able to provide a unifying framework for the analysis of topologically and morphologically altered membranes.  

Specifically, in Figure \ref{fig:Be_Sh_xi_vs_Omega}(a), we plot $\Sh$  as a function of $\Rey^\star$, and show  that appropriately rescaled data collapse reasonably well (particularly for $\Rey^\star>10^3$) with $\Sh$ increasing with $\Rey^\star$ and an inflection point for $1000<\Rey^\star<5000$. Two scaling regimes can be identified with a transition occurring at $\Rey^\star\approx 1000$:  in both regimes, $\Sh$ (i.e. $Q_{m,\infty}$) increases with $\Rey^\star$ although at different rates (with $\Sh$ increasing faster for $\Rey^\star<1000$).  The data suggests a parabolic scaling between $\Sh$ and  $\Rey^\star$ for $\Rey^\star>1000$, i.e. 
\begin{align}\label{scaling_Rey}
\Sh\sim \Rey^{\star 2} \quad \mbox{for} \quad \Rey^\star>10^3,
\end{align}
where larger inlet velocities result in larger steady-state permeate fluxes. Similarly, Figure \ref{fig:Be_Sh_xi_vs_Omega}(b) shows the relationship between $\Be$ and $\Rey^\star$ with the scaling \eqref{scaling_Rey} overlaid. As hypothesized by analogy with the benchmark case, also 
\begin{align}\label{scaling_Rey_Sh}
\Be\sim \Rey^{\star 2} \quad \mbox{for} \quad \Rey^\star>10^3,
\end{align}
i.e. larger inlet velocities result in larger overall pressure drops between the inlet and the outlet. The proposed scaling \eqref{scaling_Rey_Sh} matches the data very well.  

\begin{figure}
  \centerline{\includegraphics[width=\textwidth]{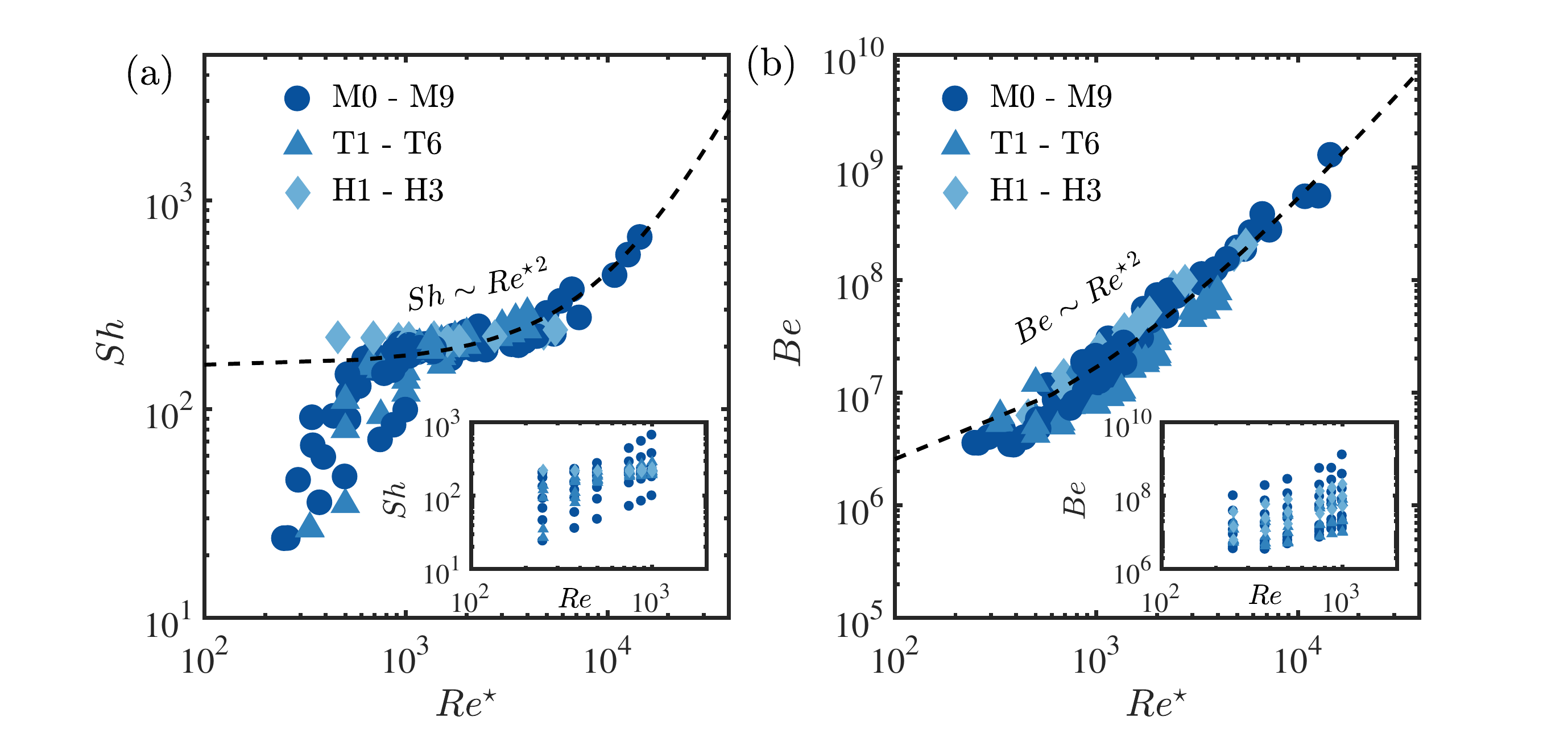}}
  \caption{(a) Sherwood number ($\Sh$) plotted as a function of the dimensionless group ($\Rey^{\star}$); (b) Bejan number ($\Be$) plotted as a function of the dimensionless group ($\Rey^{\star}$). In both plots, the dashed line is $\Be\sim \Rey^{\star 2}$.}
\label{fig:Be_Sh_xi_vs_Omega}
\end{figure}

\begin{figure}
  \centerline{\includegraphics[width=0.9\textwidth]{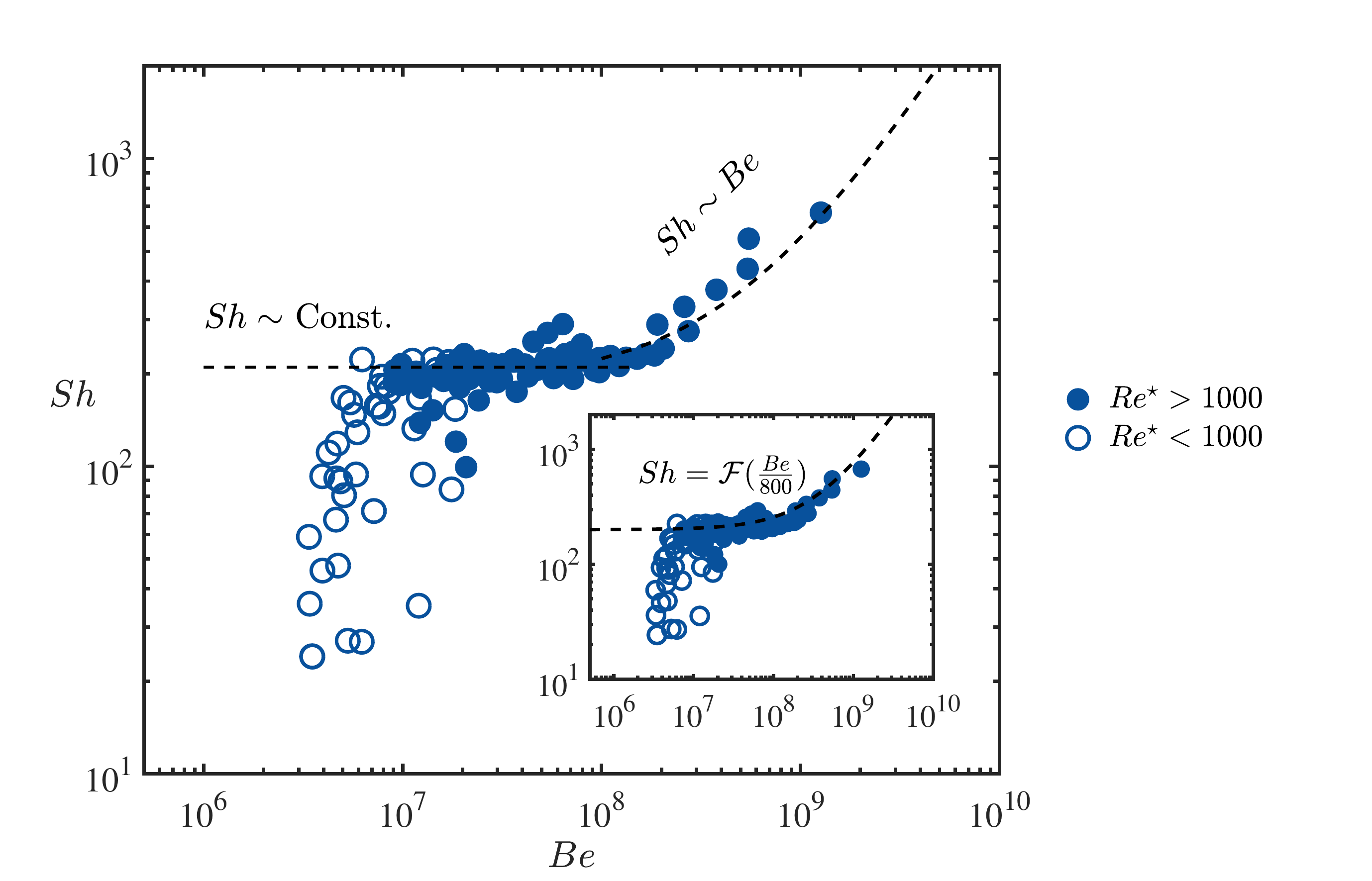}}
  \caption{$\Sh$  as a function of $\Be$ for all 114 simulations (symbols). The dashed line represents the analytical scaling of Eq.~\eqref{eqn:longtimeW_Dimless_2}. The inset shows the exact relationship needed to overlap \eqref{eqn:longtimeW_Dimless_2} with the data, where the rescaling factor is $\Be/800$.}
\label{fig:Sh_vs_Be}
\end{figure}

We now proceed by numerically validating the long-time analytical scaling relationship~\eqref{eqn:longtimeW_Dimless_2} between $\Be$  and $\Sh$. Bejam and Sherwood numbers are numerically  determined from the pressure distribution at the inlet and the permeate flux once  steady-state is reached. In Figure~\ref{fig:Sh_vs_Be}, we plot $\Sh$ as a function of $\Be$ for the 114 simulations (symbols).


Figure~\ref{fig:Sh_vs_Be} confirms the scaling relationships \eqref{scaling} derived for a rectangular membrane:  the analytical solution (dashed line in Figure~\ref{fig:Sh_vs_Be}) matches  the data points through the rescaling $\mathcal{F}(\Be/800)$ for $\Rey^\star>10^3$. For $\Rey^\star<10^3$, as Figures \ref{fig:Cb_Re_star_2D} demonstrates,  the variation of the bulk concentration $C_{\mbox{\tiny{b}}}$ with  $X$  and $Y$ is not negligible, and the assumption that $C_{\mbox{\tiny{b}}}\approx C_{\mbox{\tiny{b}}}(Z)$ is not valid any longer.


%

\subsection{Performance Index Optimization}\label{sec:optimization}


In Figure~\ref{fig:Be_Sh_xi_vs_Omega} we plot the membrane performance index for all 114 simulations: $\xi$ follows a universal nonmonotonic behavior for all types of channels (types M, T and H); first, it   increases with $\Rey^\star$ for $\Rey^\star< 10^3$, and then decreases for $\Rey^\star>10^3$. This can be explained as follows: for $\Rey^\star>10^3$, an increase in $\Rey^\star$ at a fixed $\Rey$ corresponds to an increase  of the channel waviness   $\eta$, the `roughness' height $\zeta$, or both; although these membrane alterations cause a steady increase of the permeate flux (see Figure~\ref{fig:Sh_vs_Be}(a)), this effect is out-weighted by the increase in pressure drop necessary to sustain the imposed volumetric rate  (see Figure~\ref{fig:Be_Sh_xi_vs_Omega}). As a result, $\xi$ decreases with a further increase in $\Rey^\star$, when $\Rey^\star>10^3$. Instead, for $\Rey^\star<10^3$, the increase in permeate flux  is faster than the increase in the required pressure drop, leading to a net increase of membrane performance. Importantly, Figure~\ref{fig:Be_Sh_xi_vs_Omega} shows that $\xi$ has a maximum for the values of  $\Rey^\star$ investigated, i.e. the dependence between $\xi$ and $\Rey^\star$ can be used for membrane performance optimization, both in terms of design and operating conditions.

Within each membrane type  (i.e. M or T), we select  the design that maximizes $\xi(\Rey^\star)$ across the full range of $\Rey^\star$ investigated.  Designs M4 and T1 are the best performing among the M and T designs, respectively. The overall best performing design across all categories (M, T and H) is  H1, a combination of M4 and T1, although its performance is superior to all other designs for a very limited  range of $\Rey^\star$. This demonstrates that $\xi-\Rey^\star$ curves can be used both to identify best performing designs within each class type (M or T) as  well as combine basic designs into hybrid ones to achieve improved performance. 

\begin{figure}
  \centerline{\includegraphics[width=0.9\textwidth]{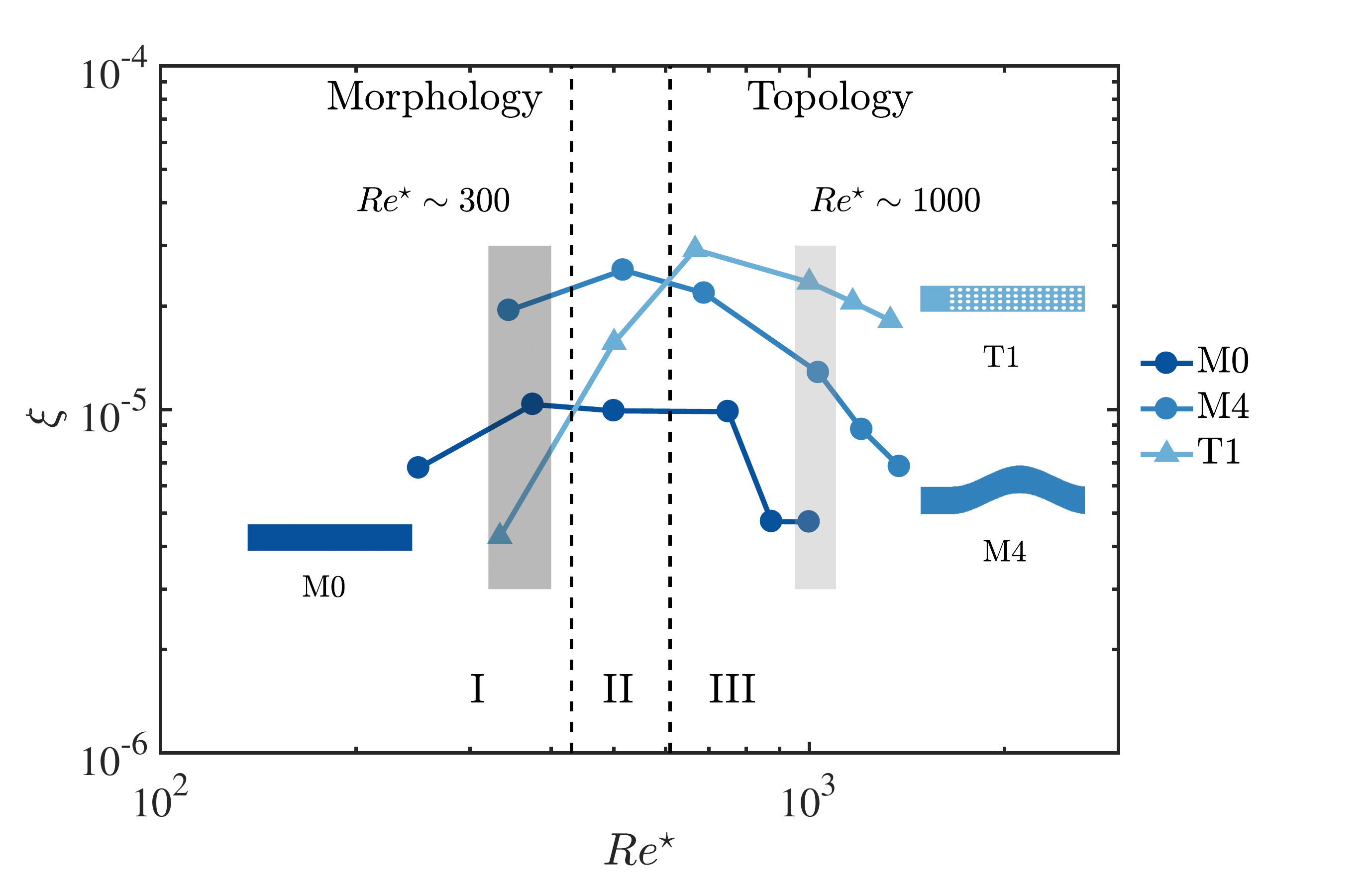}}
  \caption{Membrane performance index $\xi$ in terms of $\Rey^\star$ for the two best performing designs within their own class, M4 and T1, and the reference design M0. In region I,  morphological modifications improve $\xi$ compared to topological ones, which, instead, underperform the `do nothing' option M0. In region III, topological modifications outperforms both the benchmark design as well as the best performing M-design.}
\label{fig:Efficiency}
\end{figure}

\begin{figure}
  \centerline{\includegraphics[width=0.83\textwidth]{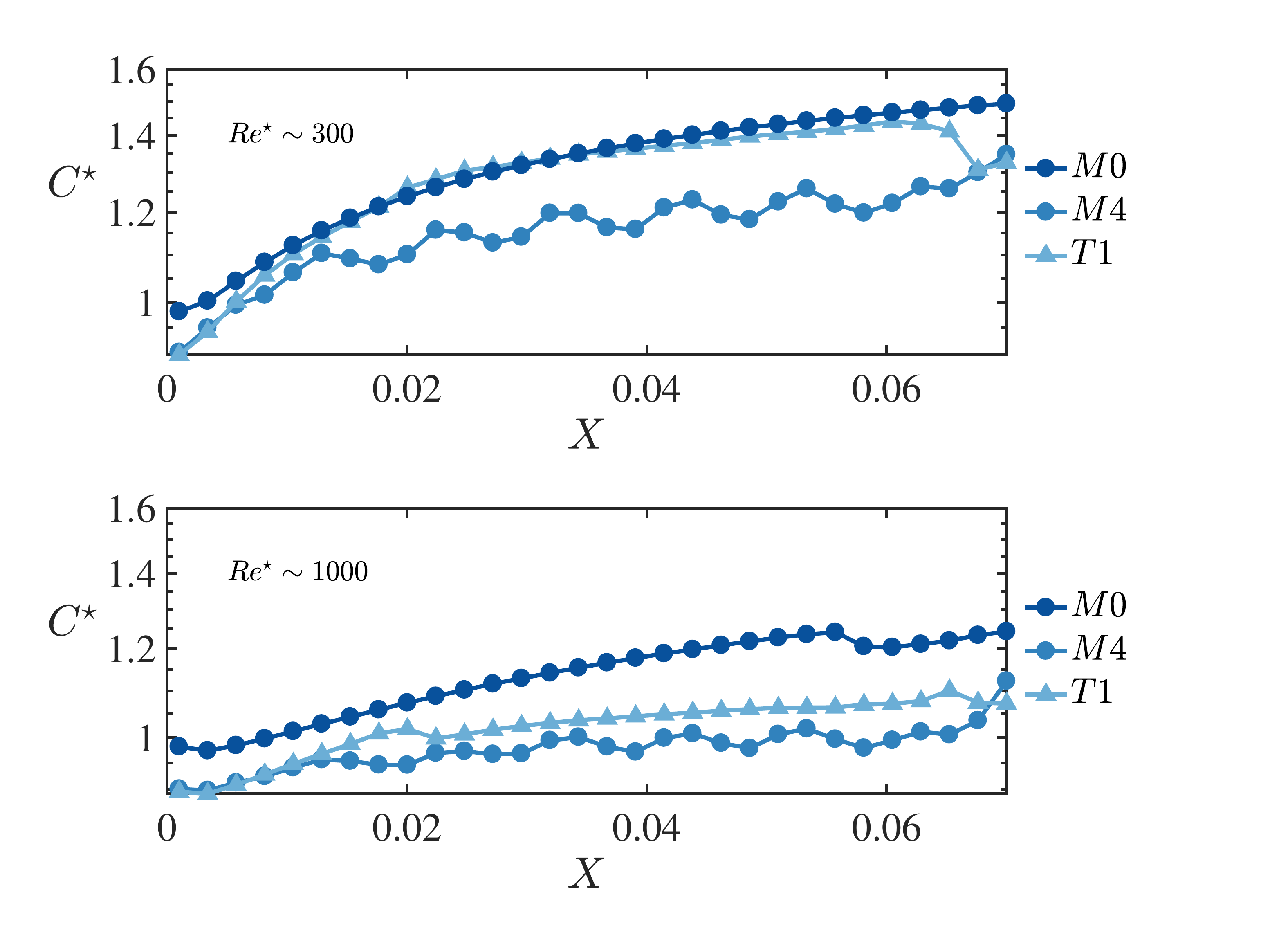}}
  \caption{Concentration polarization modulus $C^\star$, defined in Eq.~\eqref{CPM}, for M0, M4 and T1 and two values of $\Rey^\star$ ($\Rey^\star\approx 300$ top, $\Rey^\star\approx1000$ bottom).}
\label{fig:Slice_CP}
\end{figure}

Figure~\ref{fig:Efficiency} shows $\xi$ in terms of $\Rey^\star$ for M4, T1 and the reference rectangular membrane M0.  Three regions can be identified based on the magnitude of $\Rey^\star$. In Region I (i.e. at lower $\Rey^\star$) morphological alterations of the membrane improve the performance compared to the reference case M0; instead, topological modifications lead to underperformance compared to M0 (i.e. the `doing nothing' option) since the  surface pattern introduces additional roughness  and promotes foulant accumulation. In Region II, both M4 and T1 improve the system performance compared to M0,  while M4 still outperforms T1. In region III (i.e. at higher $\Rey^\star$), the trend is inverted: topological modifications maximize the membrane performance compared to both M0 and M4 since at higher $\Rey^\star$ (or velocity), surface modifications promote high permeate flux (due to an  increase of local shear stress on the membrane and a concurrent decrease in foulant accumulation) while operating at a lower  pressure drop compared to the morphologically-altered channels.
\begin{figure}
  \centerline{\includegraphics[width=0.9\textwidth]{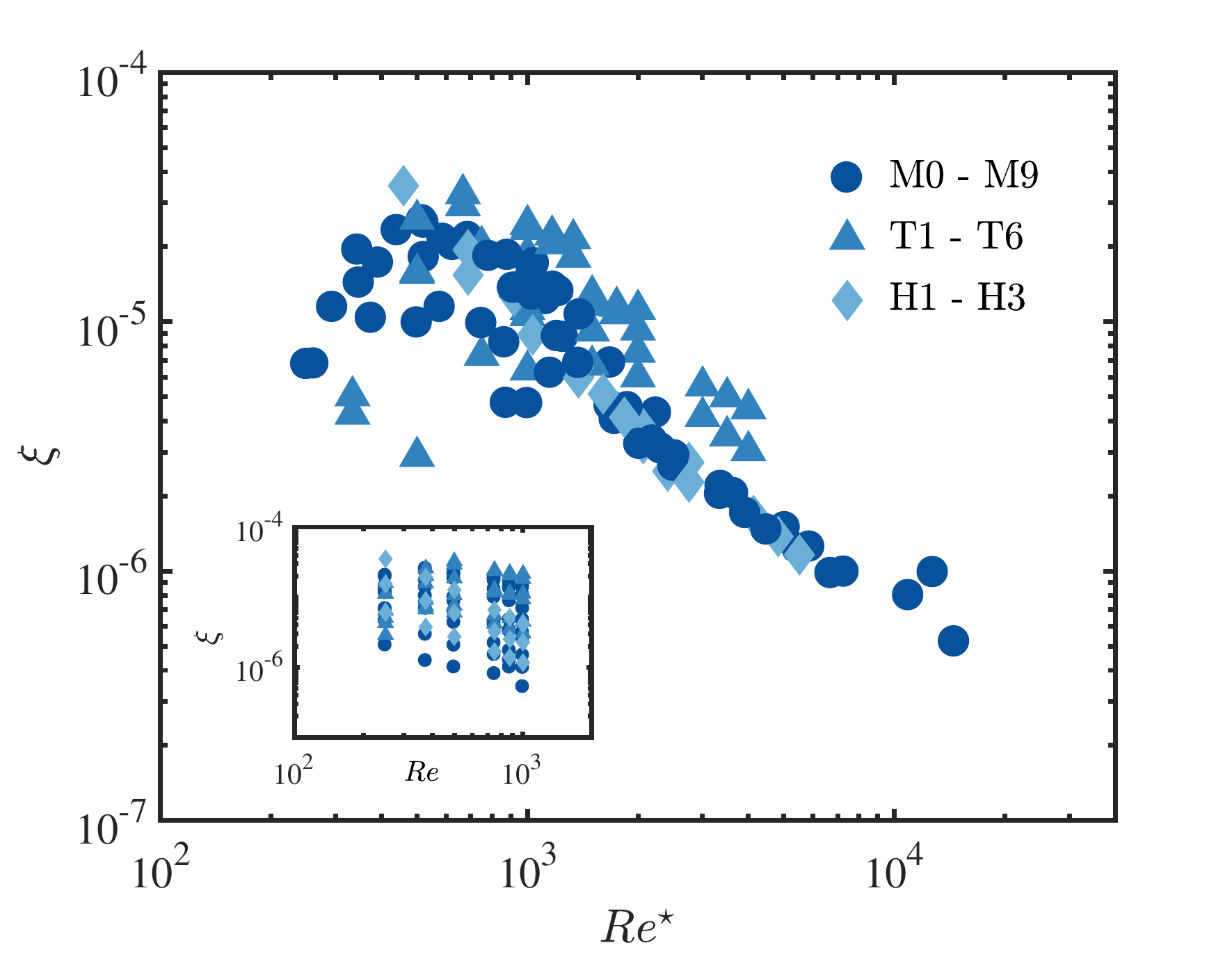}}
  \caption{Perfomance index ($\xi$) plotted as a function of the dimensionless group ($\Rey^{\star}$) for all 114 simulations.}
\label{fig:Be_Sh_xi_vs_Omega}
\end{figure}
To explore the reason that causes differences in efficiency for different patterns, in Figure \ref{fig:Slice_CP} we plot the $Y$-averaged concentration on the $Z=1.9$ mm plane as a function of $X$, i.e. the ratio between the bulk concentration evaluated in proximity of the membrane mid-plane and the inlet concentration,
\begin{align}\label{CPM}
     C^{\star}(X) = \int^{B}_{0}C_{\mbox{\tiny{b}}}\left(X,Y,Z=H^- \right) \mathrm dY,
\end{align}
also known as  concentration polarization modulus. In Figure \ref{fig:Slice_CP}, we plot the $C^{\star}(X)$  for three different geometries (M0, M4 and T1) and two different  values of $\Rey^\star$. When the $\Rey^\star<1000$, both M0 and T1 show a high concentration polarization modulus relative to M4: as a result, M4 performs better among all the cases. Additionally, since T1-design  introduces additional shear stress near the membrane surface due to its patterned surface,  it also requires higher pressure input than that needed for M0 case. This explains  why T1's performance index is lower than M0's. When the $\Rey^\star\sim 1000$, although the concentration polarization modulus of T1 and M4 is approximately the same,  M4's morphology has a much larger flow resistance (i.e. higher pressure drop requirements) which results in a lower overall performance index.

\section{Conclusions}\label{sec:conclusion}

Reverse osmosis membranes are employed in a  variety of engineering applications, ranging from waste-water purification to desalination systems. Fouling control is crucial for both efficiency enhancement and energy saving of the filtration process. Morphological (wiggliness) and topological (roughness) modifications of the membrane have been successfully employed to reduce fouling, however optimization of either  spacer morphology,  surface topology or both is still carried by trail and error. This is due to the lack of quantitative understanding of (i) the dynamic feedbacks between solute concentration in the feed solution, foulant build up on the membrane and permeate flux and (ii) the impact of morphological and/or topological modifications on membrane fouling at prescribed operating conditions. 

Here, we  develop a model, and  its corresponding customized 3D solver in OpenFOAM, that couples flow, bulk  and foulant surface concentration dynamically. The model and numerical solver are validated against experimental data of the permeate flux  conducted  by \cite{rahardianto2006diagnostic}, who studied temporal permeate flux variation and steady-state fouling pattern formation on the membrane. The code is also able to correctly predict the experimental spatial distribution of the fouling pattern. After validation, we identify  relevant dimensionless numbers involved in the problem, including the Bejam $Be$ and the Sherwood $Sh$ numbers which represent the dimensionless pressure drop along the channel and the dimensionless flux at steady state, the two primary variables to be optimized as they control directly membrane efficiency both in terms of energy consumption and generated clean water flux. We analytically derive the relationship between $Sh$ and $Be$ for a rectangular membrane and demonstrate that they exhibit the same scaling behavior in terms of Reynolds number, i.e. $Sh$ can be written as an explicit function of Bejam ($Be$), Schmidt ($Sc$) and Darcy ($Dc$) numbers, only. Two scaling behaviors are  analytically identified for $Be\rightarrow 0$ and  $Be\rightarrow \infty$ with the transition occurring at $Be^\star$. We further introduce the concept of filtration performance through the performance index $\xi$ defined as the ratio between $Sh$ and $Be$, which provides a framework to analyze the overall membrane performance both in terms of generated clean water flux and required pressure drop. The analysis derived for the benchmark rectangular membrane was then generalized to membranes with morphological and topological modifications.

Simulations conducted on 18 different designs and 6 inlet velocities, include 9 designs of membranes with sinusoidal shape (M1-M9), 6 designs of membranes patterned by cylindrical posts of different heights and arrangements (T1-T6), and 3 hybrid designs combining both morphological and topological modifications (H1-H3), in addition to the benchmark case of a classical rectangular membrane (M0), for a total of 114 simulations. The simulations reveal that  fouling  in topological or morphological altered membranes is greatly impacted by the inlet velocity, i.e. Reynolds number, with  T-type membranes better performing at high inlet velocities and M-type membranes outperforming both the benchmark M0 and T-configurations at low inlet velocities.
Since the classical Reynolds (based on the channel width length scale $B$) does not allow one to account for the additional length scales introduced by the membrane patterns or  sinusoidal shape,  in Eq.~\eqref{eq:Re_star} we introduce a modified Reynolds number, $\Rey^\star=\eta^{-2}\zeta^{-1}\Rey$, where $\eta$ ($0<\eta \leq1$) and $\zeta$ ($0\leq\zeta\leq1$) provide a measurement of the `waviness' of the channel and of the `roughness' of the membrane surface, with $\eta=1$ and $\zeta=1$ for a straight channel and a topologically unaltered membrane, respectively. The modified Reynolds number allows one to quantitatively compare the performance of membranes with different morphological and topological features under a unified framework, with $\Rey^\star$ the primary decision variable as it concurrently prescribes inlet velocity/volumetric flux and membrane geometry. Numerical simulations show that $\Rey^\star$ represents an appropriate scaling variable since the calculated $Be$, $Sh$ and $\xi$ for all 114 scenarios collapse onto universal curves, especially for $\Rey^\star>1000$, while for $\Rey^\star<1000$ the universal scaling deteriorates particularly for $Sh=f(\Rey^\star)$. 

Within this framework, we test the validity of the analytical scalings for $Be$, $Sh$ and $\xi$ derived for the straight rectangular membrane benchmark (M0), and demonstrate that $\xi-\Rey^\star$ curves can be successfully used to both identify best performing designs within each modification type (M or T), as well as employed to combine basic designs into hybrid ones to achieve improved performance. More importantly, our study provides applicability ranges in terms of the magnitude of $\Rey^\star$ within which morphological and topological modifications improve membrane efficiency (as measured by the performance index $\xi$). We identify three separate regions. At lower $\Rey^\star$, morphological changes improve the overall membrane efficiency (by reducing fouling and increasing the clean permeate flux) over the benchmark M0 and topologically altered membranes (T-designs), while the latter underperform even with respect to M0: at lower local velocities, surface roughness decreases the local velocity in proximity of the membranes and creates ideal conditions for foulant accumulation; instead, channel waviness promotes foulant segregation to the crests and troughs of the channel, while the pressure drop required to operate the system is still in check. For intermediate values of  $\Rey^\star$, both T and M designs improve upon the benchmark, although morphological modifications still outperform (at least by a factor of 2) topological ones. For higher $\Rey^\star$, T designs are superior to all M designs, i.e. surface roughness significantly  reduces fouling while only moderately increases pressure drop; instead, in M-type membranes the gain in performance due to increased permeate flux is outweighed by the increase in pressure drop needed to maintain steady state.

To conclude, we proposed a new model  to quantitatively analyze the impact that morphological and topological membrane modifications have both on fouling and energy input, while accounting for dynamic feedback between foulant bulk and surface concentration, permeate flux and pressure drop. To the best of our knowledge, this is the first work to propose a  framework (i)  that  clearly relates (micro- and meso-scale) topological and morphological structure to system- (macro-) scale function/performance and  (ii) within which the performance of different membrane designs can be assessed and optimized, while providing guidance on the most promising alteration types (morphological or topological) in terms of operating conditions.

\section{Acknowledgment}
I. Battiato gratefully acknowledges support by the National Science Foundation (NSF) through the award number 1533874, ``DMREF: An integrated multiscale modeling and experimental approach to design fouling resistant membranes''.

\bibliographystyle{jfm}
\bibliography{LB_REFERENCE}

\end{document}